\documentclass[a4paper,11pt]{article}
\pdfoutput=1 

\usepackage{jinstpub} 

\usepackage{hyperref}
\usepackage{subfigure,dcolumn}
\usepackage[T2A,T1]{fontenc}

\usepackage{siunitx}
\usepackage{subfigure}
\usepackage{tabularx}
\usepackage{multirow}
\usepackage{float}
\usepackage{bm}
\usepackage{algorithm}
\usepackage{algpseudocode}

\newcommand{\tabincell}[2]{\begin{tabular}{@{}#1@{}}#2\end{tabular}}

\title{\boldmath Universal uncertainty estimation for nuclear detector signals with neural networks and ensemble learning}

\author[a,b]{Pengcheng Ai,}
\author[a,b,1]{Zhi Deng,\note{Corresponding author.}}
\author[a,b]{Yi Wang}
\author[a,b]{and Chendi Shen}

\affiliation[a]{Department of Engineering Physics, Tsinghua University,\\Haidian District, Beijing, 100084, P. R. China}
\affiliation[b]{Key Laboratory of Particle and Radiation Imaging (Tsinghua University), Ministry of Education,\\Haidian District, Beijing, 100084, P. R. China}

\emailAdd{dengz@mail.tsinghua.edu.cn}

\abstract{Characterizing uncertainty is a common issue in nuclear measurement and has important implications for reliable physical discovery. Traditional methods are either insufficient to cope with the heterogeneous nature of uncertainty or inadequate to perform well with unknown mathematical models. In this paper, we propose using multi-layer convolutional neural networks for empirical uncertainty estimation and feature extraction of nuclear pulse signals. This method is based on deep learning, a recent development of machine learning techniques, which learns the desired mapping function from training data and generalizes to unseen test data. Furthermore, ensemble learning is utilized to estimate the uncertainty originating from trainable parameters of the network and to improve the robustness of the whole model. To evaluate the performance of the proposed method, simulation studies, in comparison with curve fitting, investigate extensive conditions and show its universal applicability. Finally, a case study is made using data from a NICA-MPD electromagnetic calorimeter module exposed to a test beam at DESY, Germany. The uncertainty estimation method successfully detected out-of-distribution samples and also achieved good accuracy in time and energy measurements.}


\arxivnumber{2110.04975} 

\begin{document}
\maketitle
\flushbottom

\section{Introduction}

In nuclear physics instruments, front-end electronics produce electrical signals when a target particle enters the sensitive volume of the detector. Due to the charge collection in a certain period of time and dedicated shaping circuits, these signals are usually unipolar or bipolar pulses with finite duration and amplitude \cite{Landsberger2015}. Many factors, such as particle type, incident energy, incident angle and location, etc., influence the shape and intensity of the nuclear pulse signal and make it a complicated statistical process to acquire relevant information from the pulse. To calibrate the performance of the detector, test beams are used to generate controlled events and response signals are recorded by subsequent electronics. Afterwards, a feature extraction algorithm (usually a rule-based procedure) is applied to the recorded signals, and measured features are fitted to a normal distribution. Equivalently, the uncertainty of measurements is assumed to be \emph{homogeneous}, at least in certain conditions. However, in real experimental circumstances, the uncertainty varies as a result of both the fluctuation of noise residing in the measurements (\emph{aleatoric} uncertainty) and the statistical model used to understand the observations (\emph{epistemic} uncertainty)\footnote{\emph{Aleatoric} uncertainty is also called \emph{statistical} uncertainty or \emph{data} uncertainty. \emph{Epistemic} uncertainty is also called \emph{systematic} uncertainty or \emph{model} uncertainty.}. The actual uncertainty is consequently \emph{heterogeneous} and changes between different sets of measurements.

Modelling heterogeneous uncertainty is a recursive topic in statistical machine learning. In recent years, with the research of deep learning \cite{8694781} going deep, researchers begin to realize that deep neural networks are able to model uncertainty in a wide range of tasks from classification to regression \cite{DBLP:conf/nips/Lakshminarayanan17,DBLP:conf/nips/KendallG17,DBLP:conf/midl/LavesIFKO20}. In these works, the main task (either producing a discrete label or a continuous value) and the associated uncertainty are combined into a unified formulation and learned jointly. In particular, ref. \cite{DBLP:conf/nips/KendallG17} combines the aleatoric uncertainty loss function with Bayesian network approximated by Monte Carlo (MC) dropout \cite{DBLP:conf/icml/GalG16}. The estimated uncertainty works as a weight coefficient to give differentiated attentions to data and effectively temper the loss caused by problematic examples. Reference \cite{DBLP:conf/nips/Lakshminarayanan17} simplifies the training and test procedure by using an ensemble of several deep neural networks to replace the complicated Bayesian network. Much computational cost is saved while comparable results are achieved by the ensemble model. Reference \cite{DBLP:conf/midl/LavesIFKO20} further uses a sigma-scaling method to mitigate the over-fitting of uncertainty estimation. Although this method was originally devised to tackle over-fitting when training examples are insufficient, we find it can also be applied when training encounters stability issues because of inconsistency in experimental data (Section \ref{sec:exp}). As a final note, the literature above assumes that the aleatoric uncertainty is of Gaussian origin; however, the same principle can be applied to other distributions by slight modifications of the loss function.

Recently, analog-to-digital converters (ADC) with high sampling rates have been used for the readout of nuclear detectors because of their flexibility and property of information preservation \cite{FABIAN2021164750,GLADEN2020164505,Zhao2019,Liu_2020,WANG2020161224}. It also provides a great opportunity to apply novel machine learning algorithms to traditional problems in nuclear instrumentation. In ref. \cite{Ai_2019}, a neural network based on auto-encoders was used for timing and characterization of pulse signals from a photon spectrometer. It achieved significantly better results than curve fitting. The signal processing neural network was validated by a field programmable gate array \cite{Chen2020} and an application specific integrated circuit (ASIC) \cite{AI2020164420}. Furthermore in ref. \cite{ai2021neural}, the authors demonstrated that neural networks could attain timing resolution near the Cram\'er Rao lower bound in radiation detector systems. In the light of these previous contributions, it is thus worthwhile to explore the ability of neural networks to quantify heterogeneous uncertainty and to establish reliable measurements in nuclear detectors.

In this paper, we aim to develop an uncertainty estimation method commonly applicable to nuclear detector signals in a wide range of scenarios. The simulation study in Section \ref{sec:sim_study} serves the purpose of validating its universality. Furthermore, to evaluate its performance in real-world detectors, we use the signals from the electromagnetic calorimeter (ECAL) designed for the Multi Purpose Detector (MPD) at the NICA collider \cite{Shen_2019,Semenov_2020} as a case study (Section \ref{sec:exp}). The contributions of the paper are listed as follows:

\begin{itemize}
	\item We design an algorithm and associated network architectures specially tailored for nuclear detector signals to extract physical features and to estimate the corresponding uncertainty. The algorithm achieves the desired accuracy in both feature regression and uncertainty estimation.
	\item We systematically demonstrate the feasibility and superiority of the proposed method through the simulation study of a typical mathematical model, in comparison with curve fitting.
	\item Test beam data using a NICA-MPD ECAL module have been used to evaluate the actual performance of the method in real-world detectors and also the ability of out-of-distribution detection.
\end{itemize}

\section{A brief introduction to ECAL at NICA-MPD}

ECAL is a sampling calorimeter with shashlik structure made by alternate layers of lead absorber and plastic scintillator. The detailed technical specification can be found in \cite{NICA-MPD-ECAL-TDR}. The main goals of the calorimeter are particle identification jointly with other parts of MPD, measurement of the photon flux, and the effective detection of photons from primary or secondary decays for reconstruction. The commissioning of the detector with the NICA collider was scheduled for December 2021 but has been postponed until one year later (December 2022) due to COVID-19. In the first stage of collider run, ECAL is equipped with silicon photomultipliers and waveform sampling--based front-end electronics. An energy resolution of 4.5\% (62.5-\si{MSPS} commercial ADC) and a 212.4 \si{ps} time resolution (5-\si{GSPS} high-speed digitizer) were achieved when the ECAL module was irradiated by 1-5 \si{GeV} electron beams \cite{LI2020162833}. Upgrades of the readout electronics are currently under research and development. ASICs integrated with high-speed ADCs (up to several hundreds of \si{MHz}) and digital feature extraction circuits are proposed to be used in the next stage of collider run to improve energy and time resolution.

Machine learning techniques, especially neural networks accelerated by digital logic, have the potential to replace traditional fixed algorithms in the front-end electronic signal analysis of electromagnetic calorimeters. This paper describes a recent attempt to deploy neural networks for the ECAL of NICA-MPD.

\section{Methodology}

\subsection{Preliminaries of neural networks and deep learning}

\begin{figure*}[!htb]
	\centering
	\includegraphics[width=.75\hsize]{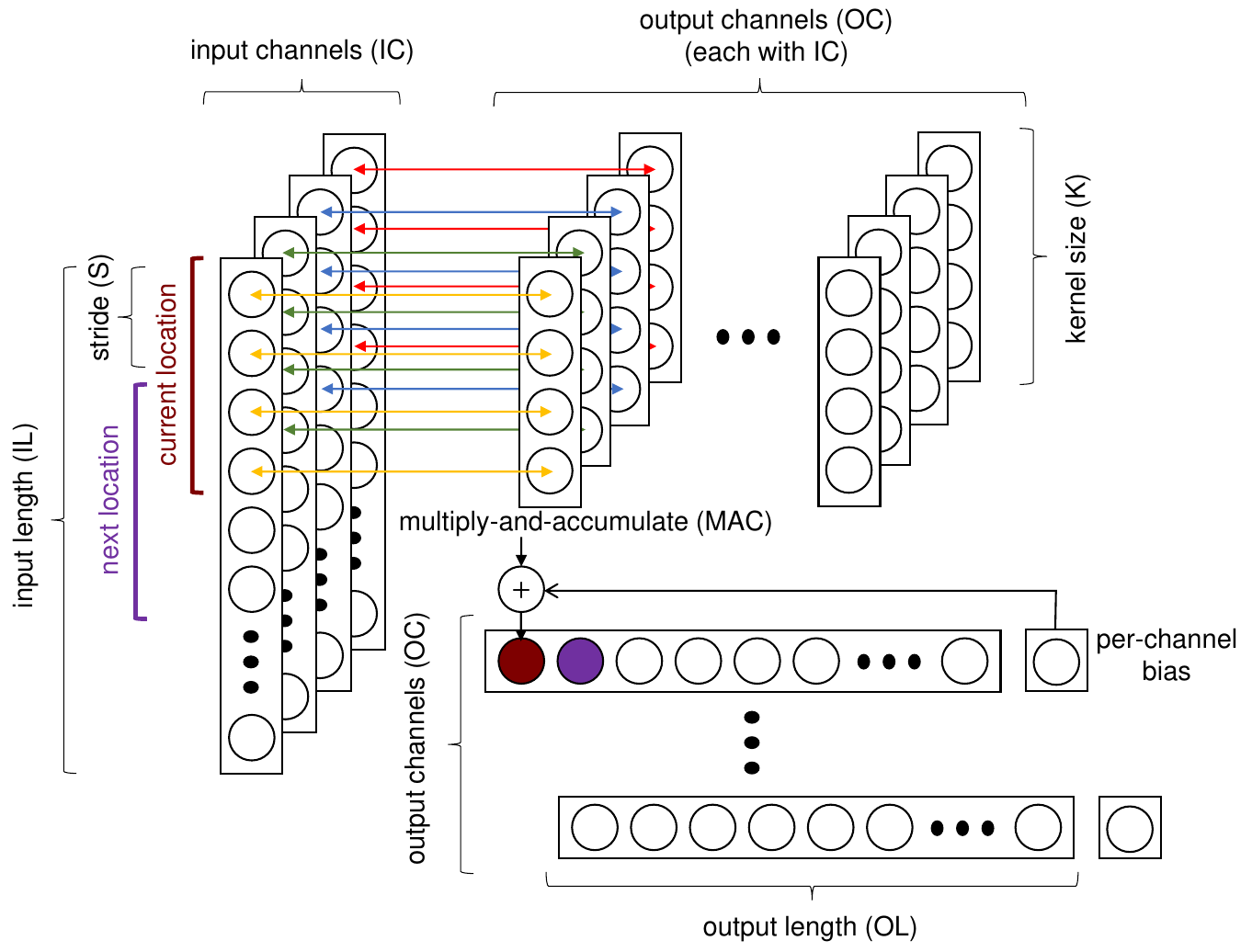}
	\caption{An illustration of one-dimensional (1D) convolution layer, the building block of 1D convolutional neural network. The optional activation after bias addition is not plotted in this functional graph.}
	\label{fig:conv1d}
\end{figure*}

Artificial neural networks were invented in the 1950's, but gained significant attention soon after AlexNet succeeded in image classification \cite{10.1145/3065386}. AlexNet was a classical convolutional neural network (CNN) aimed at two-dimensional grid data, typically in computer vision. Many well-crafted architectures \cite{DBLP:conf/cvpr/SzegedyLJSRAEVR15,DBLP:conf/cvpr/HeZRS16} were proposed after the success of AlexNet. Actually, CNNs can not only be two-dimensional, but also three-dimensional \cite{Ai_2018} or one-dimensional \cite{ABDELJABER2017154}, depending on the structure of the data.

Fig.~\ref{fig:conv1d} illustrates the working principle of one-dimensional (1D) convolution layer, which is used throughout the paper. For each slice of the kernel matrix (K*IC), it is convolved with the input feature map at a specific location to generate one line of output (OL). The convolution is in essence the multiply-and-accumulate (MAC) operation fundamental to any implementation of CNN. The convolved result is then added by the per-channel bias, and finally an optional activation function (usually the rectified linear unit, or ReLU \cite{xu2015empirical}) is applied to the sum. This process is executed recurrently to produce the OC output channels.

It should be noted that the fully-connected matrix multiplication, usually after 1D convolution, can also be represented by Fig.~\ref{fig:conv1d} when stride S equals the kernel size K. The only difference is that each slice of the kernel is used only \emph{once}, instead of reusing across the input length (IL).

\subsection{Architecture}

\begin{table}[!htb]
	\centering
	\caption{Network architecture used in the simulation. The stride (S) is kept to 2 in convolution layers, and the padding scheme is to ensure reducing the length to half.}
	\label{tab:arch_sim}
	\small
	\begin{tabularx}{0.8\hsize} { 
			>{\centering\arraybackslash}X 
			|>{\centering\arraybackslash}X 
			>{\centering\arraybackslash}X
			>{\centering\arraybackslash}X
			>{\centering\arraybackslash}X
			>{\centering\arraybackslash}X
			>{\centering\arraybackslash}X }
		\hline
		Name & IL & IC & K & OL & OC & Activation \\
		\hline
		conv1 & 32 & 1 & 4 & 16 & 16 & ReLU \\
		conv2 & 16 & 16 & 4 & 8 & 32 & ReLU \\
		conv3 & 8 & 32 & 4 & 4 & 32 & ReLU \\
		\hline
		Name & \multicolumn{2}{c}{IL*IC} & & \multicolumn{2}{c}{OL*OC} & Activation \\
		\hline
		fc1 & \multicolumn{2}{c}{4*32} & -- & \multicolumn{2}{c}{64} & ReLU \\
		fc2 & \multicolumn{2}{c}{64} & -- & \multicolumn{2}{c}{64} & ReLU \\
		fc3 & \multicolumn{2}{c}{64} & -- & \multicolumn{2}{c}{4} & None \\
		\hline
	\end{tabularx}
\end{table}

\begin{table}[!htb]
	\centering
	\caption{Network architecture used in the experiment. The stride (S) and the padding scheme are the same as those used in the simulation.}
	\label{tab:arch_exp}
	\small
	\begin{tabularx}{0.8\hsize} { 
			>{\centering\arraybackslash}X 
			|>{\centering\arraybackslash}X 
			>{\centering\arraybackslash}X
			>{\centering\arraybackslash}X
			>{\centering\arraybackslash}X
			>{\centering\arraybackslash}X
			>{\centering\arraybackslash}X }
		\hline
		Name & IL & IC & K & OL & OC & Activation \\
		\hline
		conv1 & 64 & 1 & 4 & 32 & 8 & ReLU \\
		conv2 & 32 & 8 & 4 & 16 & 16 & ReLU \\
		conv3 & 16 & 16 & 4 & 8 & 32 & ReLU \\
		conv4 & 8 & 32 & 4 & 4 & 64 & ReLU \\
		conv5 & 4 & 64 & 4 & 2 & 64 & ReLU \\
		\hline
		Name & \multicolumn{2}{c}{IL*IC} & & \multicolumn{2}{c}{OL*OC} & Activation \\
		\hline
		fc1 & \multicolumn{2}{c}{2*64} & -- & \multicolumn{2}{c}{64} & ReLU \\
		fc2 & \multicolumn{2}{c}{64} & -- & \multicolumn{2}{c}{64} & ReLU \\
		fc3 & \multicolumn{2}{c}{64} & -- & \multicolumn{2}{c}{4} & None \\
		\hline
	\end{tabularx}
\end{table}

Table \ref{tab:arch_sim} and Table \ref{tab:arch_exp} give the network architectures at the layer level used in the simulation study (Section \ref{sec:sim_study}) and experiment (Section \ref{sec:exp}). Each of the architectures is divided into two parts, the convolution layers and the regression layers. The convolution layers, without the assistance of max pooling, operate as a nonlinear filter and feature an encoder to generate the embedding of the original signal in a noisy setting. The regression layers, made up of fully-connected matrix multiplication, establish the mathematical function between the regression target and the feature embedding. In combination, the architecture is proven to work effectively for regression tasks aimed at 1D signals in nuclear detectors. The capacity of the architecture (number of layers, kernels in each layer, etc.) is carefully chosen to match the complexity of the target task and is therefore problem dependent. Some empirical research is discussed in Section \ref{sec:hyper_param_sens}, which demonstrates that the performance is not sensitive to moderate adjustment of the architecture.

\subsection{Optimization strategy}
\label{sec:opt_strategy}

When optimizing the parameters of the neural network through back-propagation, we need an optimization target, or loss function, to judge how well the network model fits the desired mapping. Assume we have $M$ regression targets, each of which needs two outputs (predictive mean and predictive variance in Equation \ref{equ:nn}). With $N$ i.i.d. examples, the loss function is given by Equation \ref{equ:loss}:

\begin{equation} \label{equ:nn}
	\bm{f}_{\mathrm{NN}} (\bm x; \bm \theta) = \left[\bm \mu(\bm x), \bm \sigma(\bm x)^2 \right], \quad \bm \mu, \bm \sigma^2 \in \mathbb{R}^M,\ \ \bm \sigma^2 > 0
\end{equation}

\begin{equation} \label{equ:loss}
	\mathcal{L}(\bm \theta) = \frac{1}{N} \sum_{i=1}^{N} \sum_{j=1}^{M} \frac{1}{2 \sigma_{j} ( \bm{x}^{(i)} )^2} \left| \left| y_j^{(i)} - \mu_j ( \bm{x}^{(i)} ) \right| \right|^2 + \frac{1}{2} \log \sigma_{j}( \bm{x}^{(i)} )^2
\end{equation}

\noindent where $\bm f$ is the mapping function of the neural network model, $\bm x$ is the input time series of signal, $\bm y$ is the ground-truth label indicating the desired output, $\bm \theta$ is the trainable parameters of the neural network, and $\bm \mu, \bm \sigma^2$ are the predictive mean and predictive variance, respectively. This loss function is derived from the log-likelihood assuming Gaussian distributions and is similar to what is used in \cite{DBLP:conf/nips/KendallG17} and \cite{DBLP:conf/midl/LavesIFKO20}. The major difference here is to separate the predictive variance for each feature: for each example $\bm{x}$, we learn a $(\mu_j, \sigma_j^2)$ pair per feature, instead of sharing $\sigma^2$ for all features.

In practice, in order to force $\sigma_j^2$ to be positive without explicit constraints, we let the network to output $\log(\sigma_j^2)$ and use the exponential function when computing the loss. This ensures the predictive variance is always positive. Because both the time and energy are desired simultaneously, we use $M$ = 2 and design the output length of the neural network to be 4 (Table \ref{tab:arch_sim} and Table \ref{tab:arch_exp}).

\subsection{Ensemble learning}

\begin{figure*}[!htb]
	\centering
	\includegraphics[width=.55\hsize]{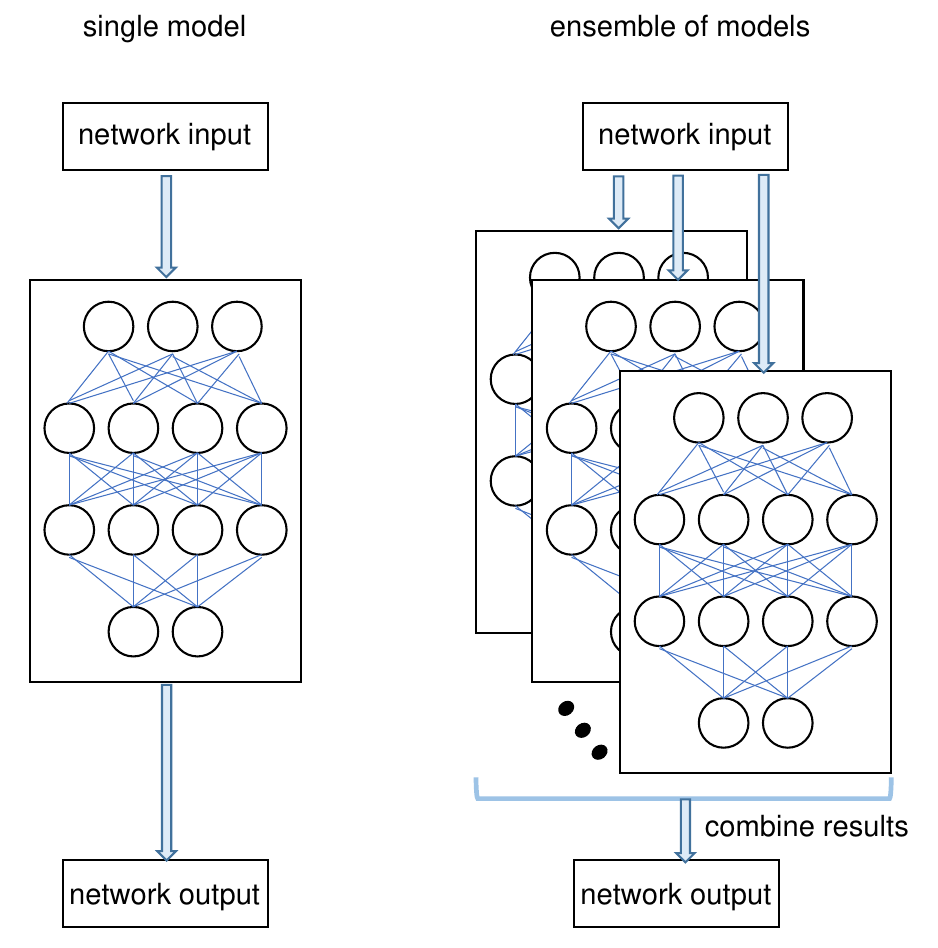}
	\caption{The workflow of a single neural network model, and the method to combine several individual neural network models into an ensemble in the test stage. The inner structure in the square, which can be any network architecture in practice, is only for demonstration.}
	\label{fig:nn_ensemble}
\end{figure*}

A single neural network is sufficient to model the aleatoric uncertainty due to the measurement noise. However, it is unable to describe the epistemic uncertainty regarding how well our neural network model understands the data. To model epistemic uncertainty, an ensemble considering the variations of network parameters is needed. Besides, ensemble learning combines the individual results of several weak models and thus improves the reliability when generalizing to test data. The formation of the ensemble from individual neural networks is shown in Fig. \ref{fig:nn_ensemble}. In the training stage, several independent neural networks are initialized with different parameters and trained separately. In the test stage, the same input data is fed to each neural network in the ensemble, and the results are combined at the other end. The comprehensive mean and variance are determined as follows:

\begin{equation}
	\mu_{j, *}(\bm x) = \frac{1}{T} \sum_{t=1}^{T} \mu_{j, \bm{\theta}_t}(\bm x)
\end{equation}

\begin{equation}
	\sigma_{j, *}(\bm x)^2 = \frac{1}{T} \sum_{t=1}^{T} \left( \sigma_{j, \bm{\theta}_t}(\bm x)^2 + \mu_{j, \bm{\theta}_t}(\bm x)^2 \right) - \mu_{j, *}(\bm x)^2
\end{equation}

\noindent where $T$ is the number of individual models, $\bm \theta_t$ is the trainable parameters of the t-th model, $\mu_{j, \bm{\theta}_t}, \sigma_{j, \bm{\theta}_t}^2$ are mean and variance outputs for the j-th regression target in the t-th model, and $\mu_{j, *}, \sigma_{j, *}^2$ are the combined mean and variance for the j-th regression target.

For the predictive mean, the introduction of ensemble learning can improve the robustness to over-fitting and make the model generalize well to unseen examples. For the predictive variance, it is an effective way to quantize epistemic uncertainty by exploring variations of the model parameters.

\section{Simulation Study}
\label{sec:sim_study}

In this section, we study the possibilities and advantages of neural networks and ensemble learning to characterize the measurement uncertainty of nuclear detector signals. This is done by comparing neural networks to nonlinear least squares curve fitting when the signals have a precise mathematical function. In certain conditions (uncorrelated noise), curve fitting gives the near-optimal estimation of uncertainty; in other conditions (correlated noise), curve fitting is sub-optimal and unable to estimate the uncertainty precisely. By observing the behaviors of neural networks and curve fitting in different conditions, we can gain both qualitative and quantitative understandings of the proposed method.

We use the following CRRC waveform (generated by step function passing through a CRRC circuit) in the simulation study:

\begin{equation}
	s(t) = K \left( \frac{t - t_0}{\tau} \right) e ^ {- (t-t_0)/\tau} u(t-t_0)
\end{equation}

\begin{figure*}[!htb]
	\centering
	\includegraphics[width=.55\hsize]{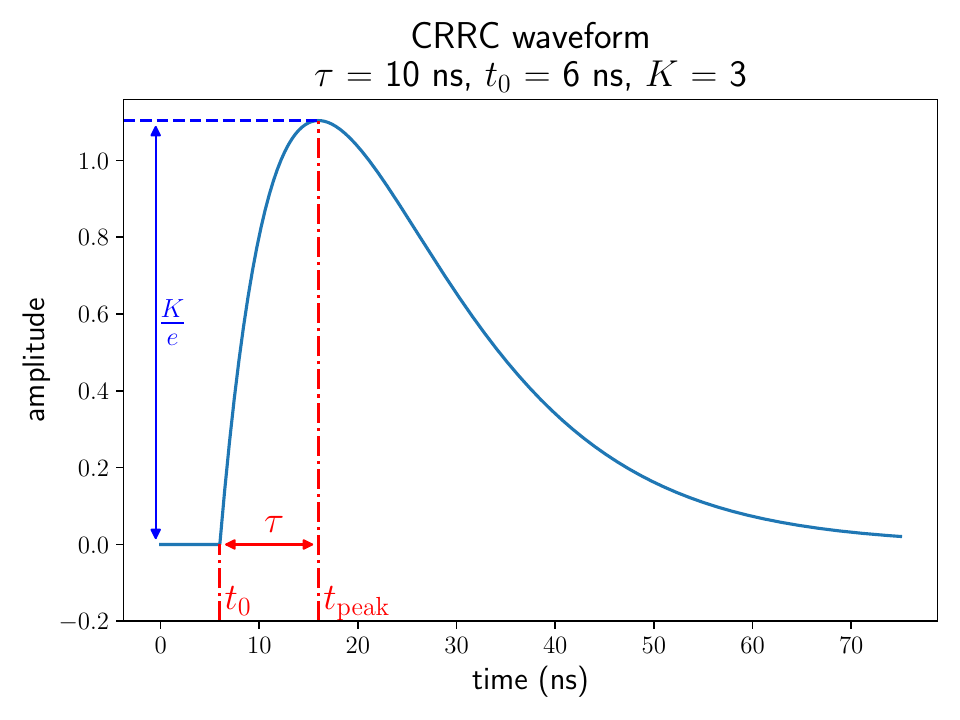}
	\caption{An example of the CRRC waveform with annotations.}
	\label{fig:sim_crrc_waveform}
\end{figure*}

\noindent where $\tau$ is the shape coefficient related to the resistance and capacitance in the CRRC circuit, $t_0$ is the start time of the waveform (timing label), $K$ is the amplitude of the waveform (energy label), and $u(t)$ is the step function. Fig. \ref{fig:sim_crrc_waveform} gives an example of the CRRC waveform with typical parameters. The waveform is sampled at a fixed interval to produce the input time series. Throughout the simulation, we use $\tau$ = 10 \si{ns}, and sample 32 points with 2 \si{ns} interval (500 \si{MSPS}). When generating the datasets, $K$ is sampled from a uniform distribution between 2.0 and 4.0, and $t_0$ is sampled from a uniform distribution between 0 \si{ns} and 4 \si{ns} (with 6 \si{ns} offset).

When training the ensemble of neural networks, we use 8000 examples in the training set and 2000 examples in the test set. With the batch size of 64, we train for 40 epochs using the Adam \cite{DBLP:journals/corr/KingmaB14} optimization algorithm at 0.001 learning rate to minimize the loss in Section \ref{sec:opt_strategy}. We follow \cite{DBLP:conf/nips/Lakshminarayanan17} and use an ensemble of 5 neural networks\footnote{In practice, increasing the ensemble size can potentially improve the performance; however, the improvement will saturate when the size is above a certain number \cite{DBLP:conf/iclr/AshukhaLMV20}. Here, we choose 5 as a balance of performance requirements and computational cost.} throughout the paper. We also experiment with Monte Carlo (MC) dropout \cite{DBLP:conf/icml/GalG16,DBLP:conf/nips/KendallG17} (a Bayesian network based method) for comparison. For MC dropout, the batch size is 256, the number of training epochs is 160, the dropout rate is 0.01, and the number of forward passes with dropout at test time is 100. The software is implemented with Keras \cite{chollet2015}, an open-source deep learning framework, on a desktop computer with RTX 2060 Super GPU (8 \si{GB} video memory).

\subsection{Predictive performance}

\begin{table*}[!htb]
	\centering
	\caption{Quantitative results of predictive performance for different conditions in the simulation. The bold number in each column of the table section (separated by simulation conditions) indicates the best result of different methods.}
	\label{tab:sim_num}
	\fontsize{7.5pt}{12pt}\selectfont
	\setlength{\tabcolsep}{3pt}
	\begin{tabular}{c|c|ccccc|ccccc}
		\hline
		\multicolumn{2}{c|}{\multirow{2}{*}{}} & \multicolumn{5}{c|}{time} & \multicolumn{5}{c}{energy} \\
		\cline{3-12}
		\multicolumn{2}{c|}{ } & NLL & B-UCE & A-UCE & bias (\si{ns}) & precision (\si{ns}) & NLL & B-UCE & A-UCE & bias & precision \\
		\hline
		\multirow{3}{*}{\tabincell{c}{unimodal\\uncorrelated}} & fit & -4.243e-1 & \textbf{2.542e-3} & \textbf{1.547e-3} & 0.007 & 0.166 & -1.681 & 4.179e-4 & \textbf{9.110e-4} & \textbf{0.013\%} & 1.552\% \\
		\cline{2-12}
		& MC drop. & -3.703e-1 & 1.151e-2 & 2.032e-2 & 0.017 & 0.168 & -1.611 & 1.401e-3 & 1.984e-2 & 0.150\% & 1.610\% \\
		\cline{2-12}
		& ensem. & \textbf{-4.318e-1} & 5.559e-3 & 3.233e-3 & \textbf{-0.003} & \textbf{0.163} & \textbf{-1.697} & \textbf{3.776e-4} & 4.505e-3 & -0.083\% & \textbf{1.535\%} \\
		\hline
		\multirow{3}{*}{\tabincell{c}{multimodal\\uncorrelated}} & fit & 1.800e-1 & 3.244e-2 & \textbf{2.233e-3} & 0.011 & 0.390 & -1.133 & 2.329e-3 & \textbf{4.885e-4} & \textbf{0.082\%} & 3.485\% \\
		\cline{2-12}
		& MC drop. & 1.475e-1 & \textbf{2.370e-2} & 1.629e-2 & -0.027 & 0.348 & -1.055 & 2.464e-3 & 5.948e-2 & 0.715\% & 3.418\% \\
		\cline{2-12}
		& ensem. & \textbf{9.253e-2} & 2.417e-2 & 4.291e-3 & \textbf{0.005} & \textbf{0.342} & \textbf{-1.148} & \textbf{2.015e-3} & 3.637e-3 & 0.117\% & \textbf{3.329\%} \\
		\hline
		\multirow{3}{*}{\tabincell{c}{unimodal\\correlated}} & fit & 8.113e-1 & 5.304e-2 & 9.149e-2 & \textbf{0.001} & 0.277 & -6.417e-1 & 3.461e-3 & 8.731e-2 & 0.043\% & 2.528\% \\
		\cline{2-12}
		& MC drop. & -2.947e-1 & 1.433e-2 & 1.854e-2 & -0.015 & \textbf{0.183} & -1.278 & 2.273e-3 & 1.389e-2 & 0.175\% & 2.271\% \\
		\cline{2-12}
		& ensem. & \textbf{-3.097e-1} & \textbf{5.877e-3} & \textbf{9.465e-3} & -0.009 & 0.185 & \textbf{-1.338} & \textbf{6.154e-4} & \textbf{1.900e-3} & \textbf{-0.024\%} & \textbf{2.196\%} \\
		\hline
		\multirow{3}{*}{\tabincell{c}{multimodal\\correlated}} & fit & 1.441 & 3.242e-1 & 9.060e-2 & 0.022 & 0.665 & -1.301e-1 & 1.697e-2 & 7.239e-2 & 0.273\% & 5.624\% \\
		\cline{2-12}
		& MC drop. & 2.708e-1 & 3.190e-2 & \textbf{9.795e-4} & \textbf{0.002} & 0.413 & -7.628e-1 & 5.075e-3 & 1.270e-2 & \textbf{-0.095\%} & 4.628\% \\
		\cline{2-12}
		& ensem. & \textbf{2.254e-1} & \textbf{2.166e-2} & 1.584e-3 & 0.012 & \textbf{0.402} & \textbf{-8.060e-1} & \textbf{3.168e-3} & \textbf{6.299e-3} & 0.474\% & \textbf{4.600\%} \\
		\hline
	\end{tabular}
\end{table*}

\begin{figure*}[!htb]
	\centering
	\includegraphics[width=.65\hsize]{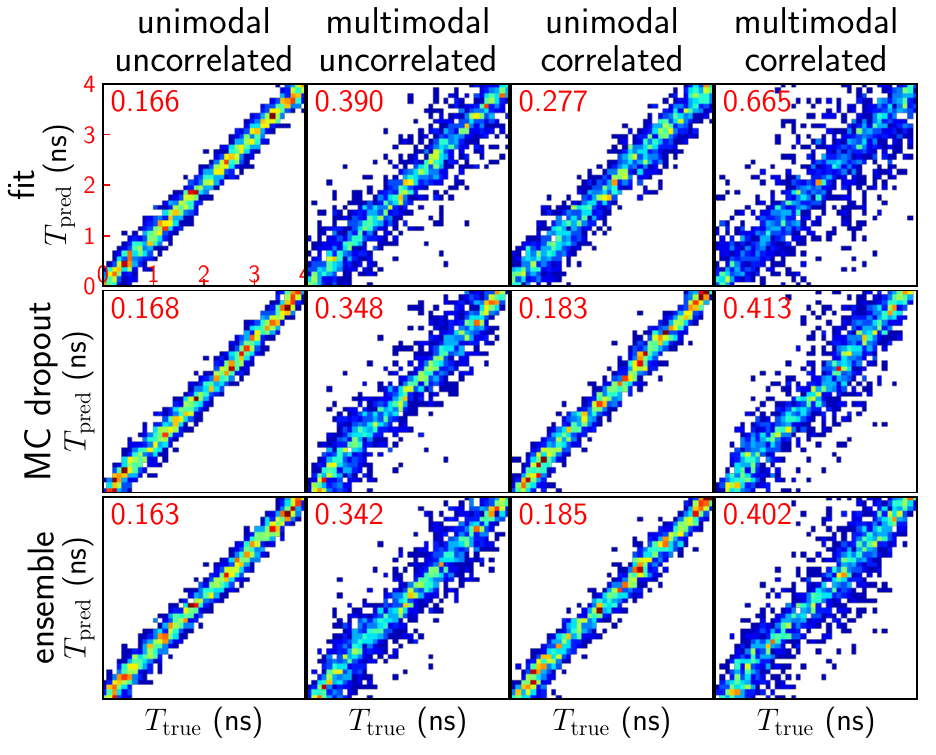}
	\includegraphics[width=.65\hsize]{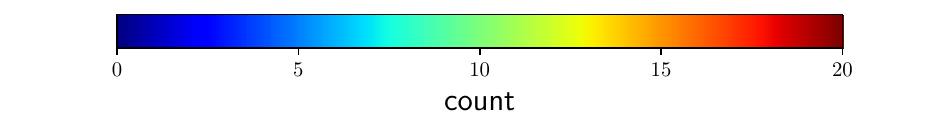}
	\caption{Two-dimensional histograms show predicted time vs. ground-truth time using three different methods (curve fitting, MC dropout and ensemble) in four conditions. The red numbers in the top left-hand corner of each box indicate timing precision in \si{ns}. The axis scales are identical in each case, and are shown in the top left-hand box.}
	\label{fig:sim_pred_value_main}
\end{figure*}

\begin{figure*}[!htb]
	\centering
	\includegraphics[width=.85\hsize]{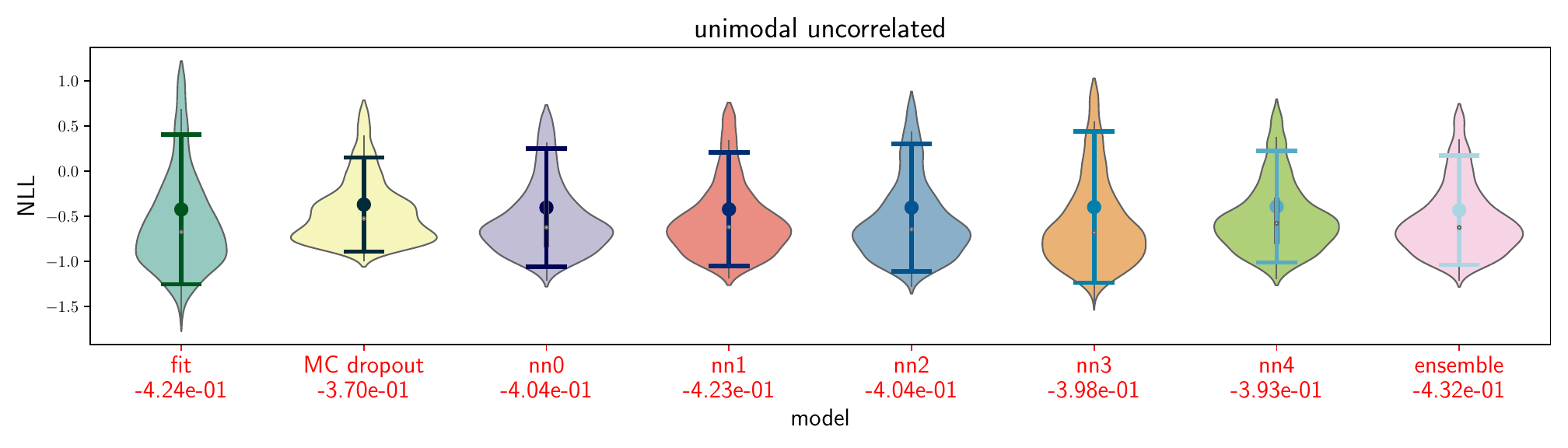}
	\includegraphics[width=.85\hsize]{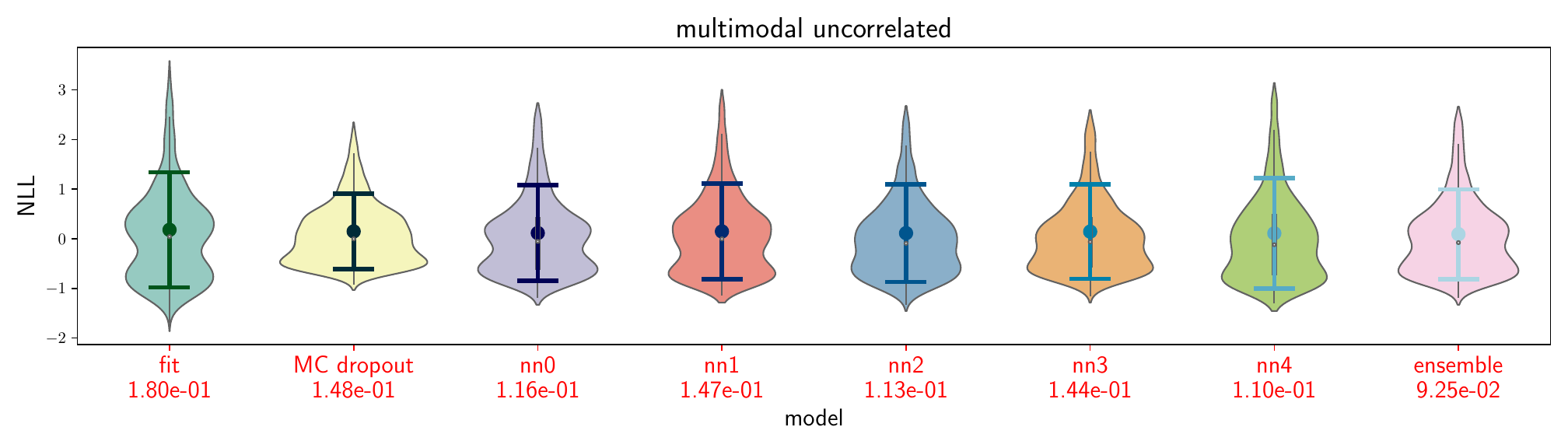}
	\includegraphics[width=.85\hsize]{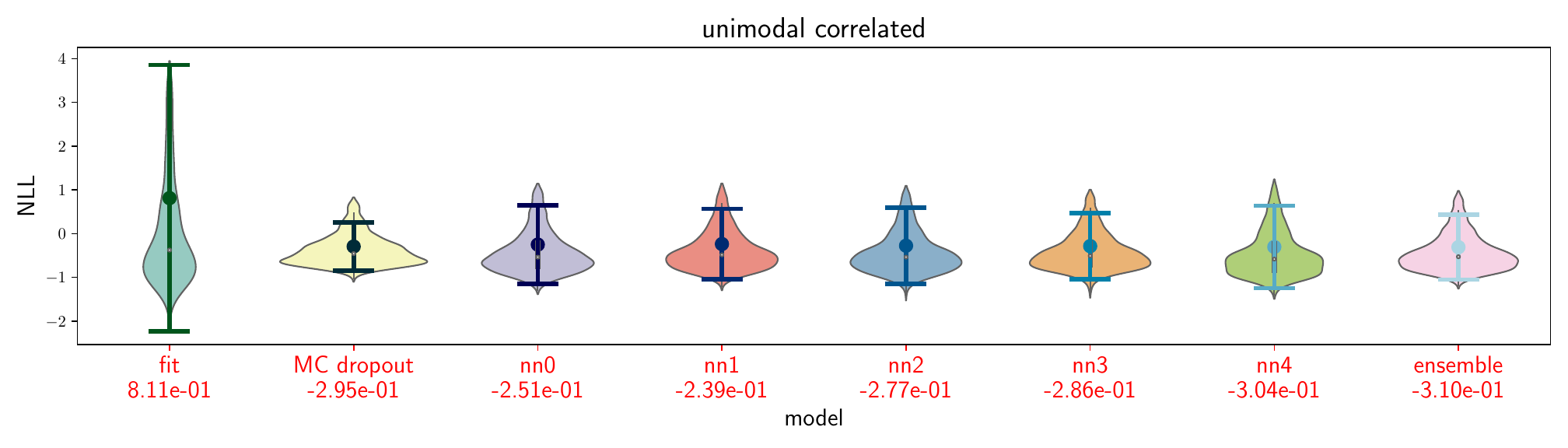}
	\includegraphics[width=.85\hsize]{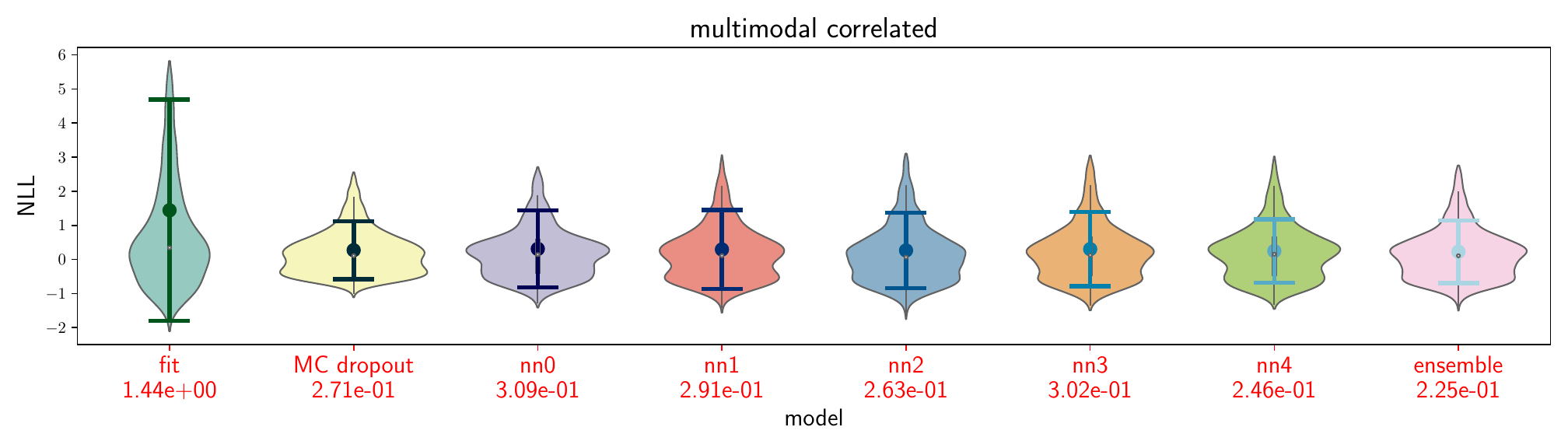}
	\caption{Categorical plots show the distributions of NLL values using three different methods (for ensemble, each single neural network is also shown). To draw violin plots, a few outliers (significantly larger) are excluded for better visual results. The overlying error bars use all examples including outliers. The red numbers below the x-axis tick labels indicate average NLL values.}
	\label{fig:sim_violin}
\end{figure*}

In Table \ref{tab:sim_num}, we list negative log-likelihood (NLL) \cite{DBLP:conf/nips/Lakshminarayanan17}, binned uncertainty calibration error (B-UCE, the UCE defined in \cite{DBLP:conf/midl/LavesIFKO20}), accumulated uncertainty calibration error (A-UCE, the calibration error for regression defined in \cite{DBLP:conf/icml/KuleshovFE18}), and the bias and precision for the time and energy predictions in each condition. All these metrics are explained in Algorithm \ref{alg:metrics} of the appendix. In them, NLL on the test set is a comprehensive index to judge how well the model estimates the true value and assigns uncertainty to each estimation, B-UCE/A-UCE measures the results by integrating the predictive variance and the actual error, and bias/precision mainly deals with the predictive mean. The smaller values of these measures, the better, except for the bias which needs a small absolute value.

Fig. \ref{fig:sim_pred_value_main} (as well as Fig. \ref{fig:sim_pred_value_supp} in the appendix) gives the performance of time prediction with three methods in four different conditions. In the varying range of the ground-truth time, a good prediction will show high linearity and little spread on the two-dimensional histogram. It can be seen that MC dropout and ensemble of neural networks are better than curve fitting when the noise distribution is multimodal or correlated. In total, ensemble of neural networks achieves the best precision in three out of four conditions.

Fig. \ref{fig:sim_violin} visualizes the distribution of NLL values by each method in each condition, in the time section. To minimize the loss in Section \ref{sec:opt_strategy}, the model tends to shrink uncertainty values when predictions are accurate, so that the Gaussian distributions in the log-likelihood are steep; however, if there are examples with large prediction errors, they will have even smaller log-likelihood values with steeper distributions so as to counteract the former efforts and to increase the loss. In view of this effect, NLL is better at quantifying the performance of predictive variance than least squares or other measures. In Fig. \ref{fig:sim_violin}, it can be seen that ensemble of neural networks is always the best (smallest average NLL) in all conditions. Besides, methods based on neural networks show two peaks in multimodal conditions. Since the synthetic data have a simple generative distribution, the aleatoric uncertainty dominates. Therefore, the improvement of ensemble is not very prominent compared with individual neural networks. Later in Section \ref{sec:out-of-dist}, we will display the advantage of ensemble learning in more detail.

Another comment is about MC dropout, which needs many more forward passes than ensemble of neural networks to achieve similar performance. Usually, this is not feasible if we want to deploy the algorithm onto hardware (Section \ref{sec:quan}). Since there are no fundamental differences between MC dropout and ensemble of neural networks, we will use the latter, the proposed method in this paper, in the following sections.

\subsection{Unimodal uncorrelated condition}

\begin{figure*}[!htb]
	\centering
	\subfigure[uncorrelated]{
		\label{fig:sim_norm_calibre_uncorr}
		\includegraphics[width=.48\hsize]{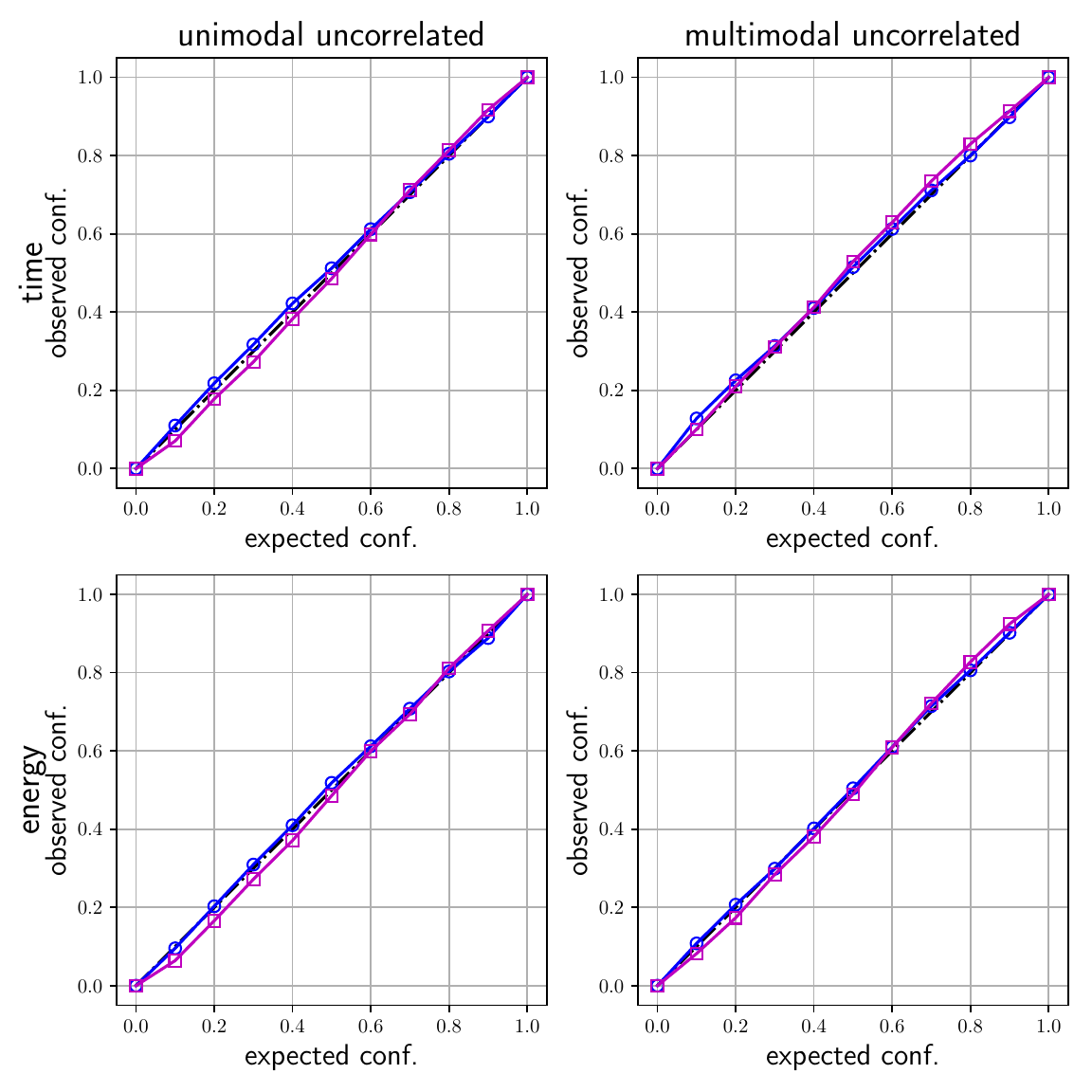}
	}
	\subfigure[correlated]{
		\label{fig:sim_norm_calibre_corr}
		\includegraphics[width=.48\hsize]{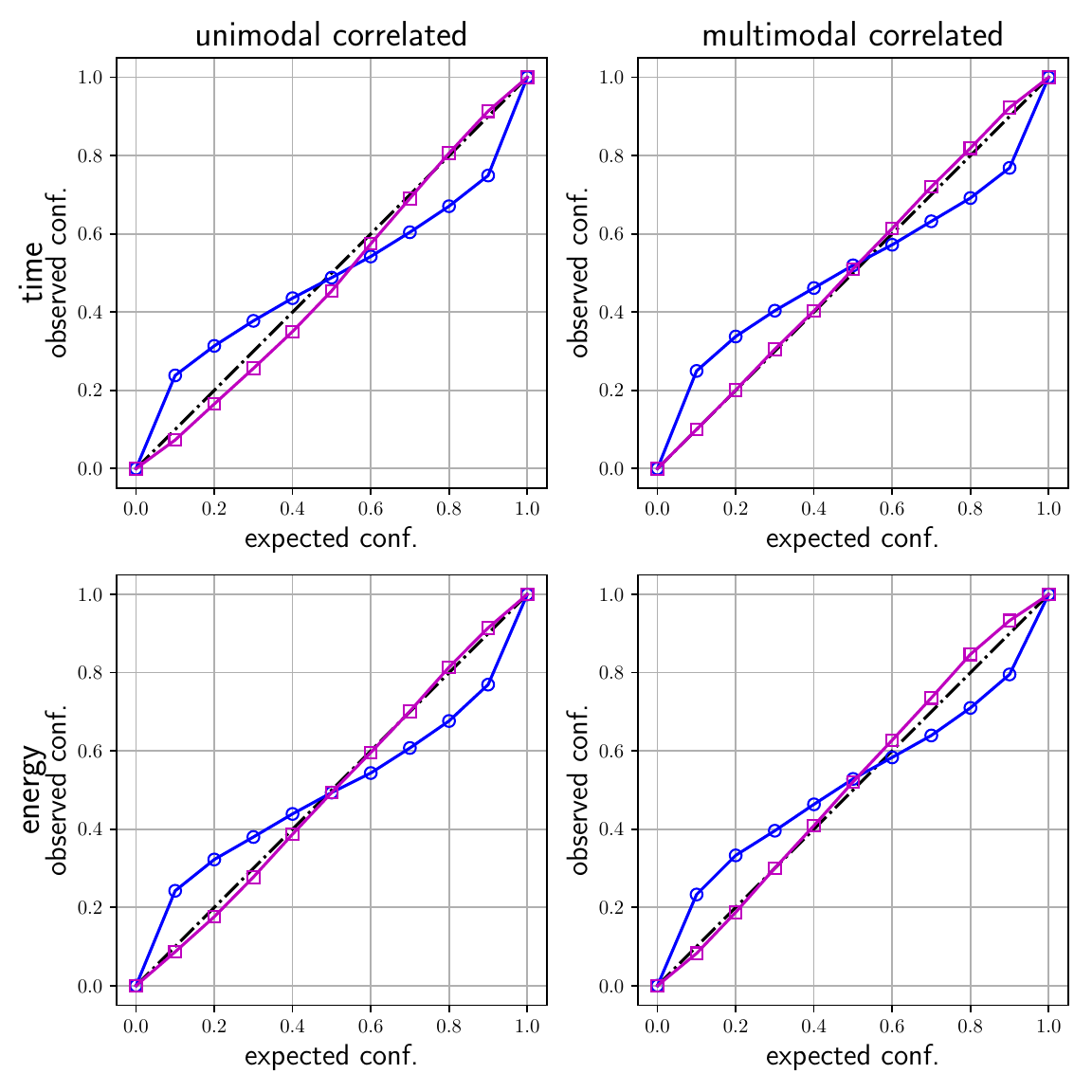}
	}
	\caption{Normalized calibration plots for (a) uncorrelated and (b) correlated conditions in the simulation. The blue lines with circle markers represent curve fitting, and the magenta lines with square markers represent ensemble of neural networks.}
\end{figure*}

In this setting, a random Gaussian white noise with 0.05 standard deviation is added to the sampling points. Since noises at different times are completely uncorrelated, nonlinear least squares curve fitting gives the maximum likelihood estimation \cite{Ai_2019} with reasonable standard errors of fitting parameters. It should be noted that this condition is a strong assumption and very idealized since the underlying mathematical function is unknown and variable in reality.

To assess the overall quality of predictive uncertainty, we draw the normalized calibration plots \cite{DBLP:conf/icml/KuleshovFE18} to visualize observed confidence levels at different expected confidence levels (refer to Algorithm \ref{alg:metrics} in the appendix). When the uncertainty is perfectly calibrated, the above two confidence levels should match at any interval, resulting in a straight line from (0, 0) to (1, 1). In the first column of Fig. \ref{fig:sim_norm_calibre_uncorr}, it can be seen that both curve fitting and ensemble of neural networks achieve good calibration, while curve fitting is slightly better, in the unimodal uncorrelated condition. This is further proved by the A-UCE column in Table \ref{tab:sim_num}, which is a direct measure of the deviation to the straight line in the normalized calibration plot.

The first three rows of the table compare three methods in this condition. In general, ensemble of neural networks performs at least as well as curve fitting (reflected by NLL and precision). The reason why ensemble can be even better is that curve fitting may stop at a position which is not the global minimum during the numerical optimization process. In contrast, ensemble of neural networks is more robust given sufficient training examples.

\subsection{Correlation}

Correlation is pervasive in nuclear detector signals because of dependence between sampling points. To study this condition, a random Gaussian white noise with 4.0 standard deviation is injected at the source side before the CRRC circuit. After filtering by the bandpass CRRC shaper, the noise is significantly attenuated and rendered a low-frequency behavior. Because of correlation, curve fitting is no longer the maximum likelihood estimation so that it may give sub-optimal fitting results and unreasonable standard errors. However, because of the highly nonlinear mapping, neural networks are able to work well even though the input data is correlated \cite{Ai_2019}.

In Fig. \ref{fig:sim_norm_calibre_corr}, it can be seen that the normalized calibration plots of curve fitting apparently deviate from the straight line. In comparison, ensemble of neural networks still keeps to the straight line just as it does in the former condition. This shows the advantage of neural networks and ensemble learning when the statistical distributions of input data are not independent.

The last six rows of Table \ref{tab:sim_num} give quantitative results in the correlated setting. Ensemble of neural networks is consistently and significantly better than curve fitting in all measures except for the bias which is not a serious issue in this problem\footnote{In large detector systems, bias is common due to the mismatch between detectors and recorded data and will be systematically corrected in the calibration process.}. Among them, large improvements of NLL and precision are observed. These further demonstrate the advantage of the proposed method.

\subsection{Multi-modality}

\begin{figure*}[!htb]
	\centering
	\subfigure{
		\includegraphics[width=.75\hsize]{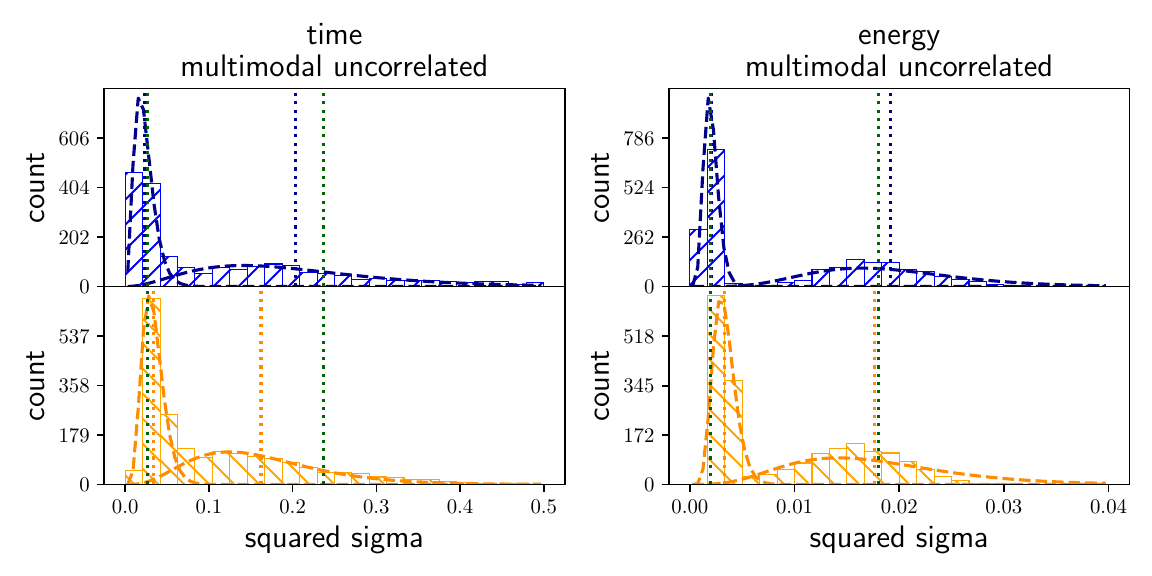}
	}
	\subfigure{
		\includegraphics[width=.75\hsize]{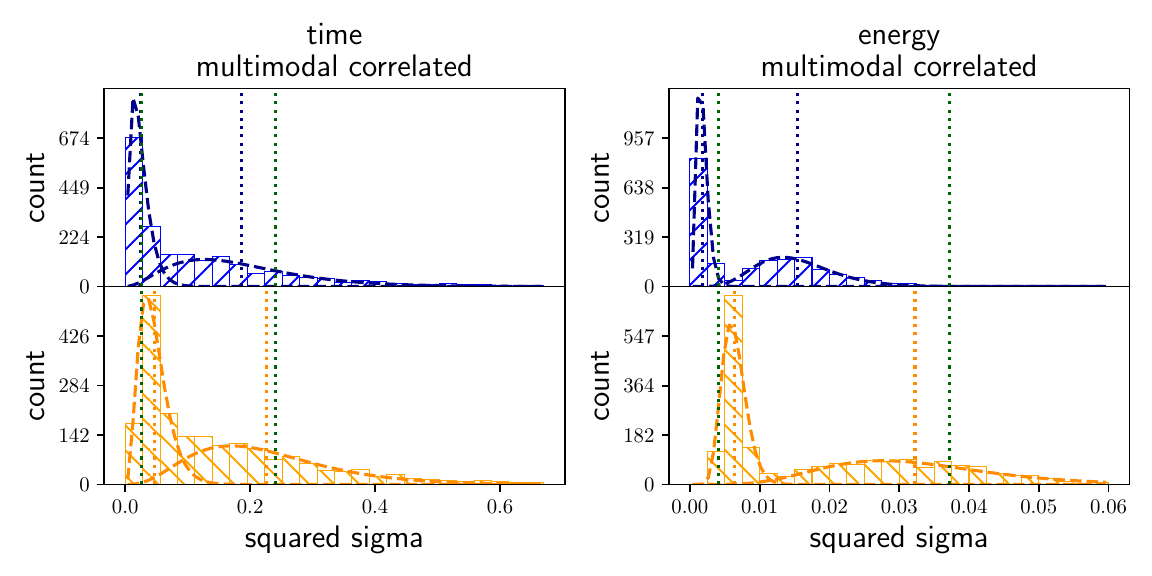}
	}
	\caption{Distribution of uncertainty predictions for curve fitting (blue, diagonal) and ensemble of neural networks (orange, back diagonal), in the multimodal conditions. We fit each of the histograms to a mixture of two chi-squared distributions. The blue/orange dotted lines represent the mean values of chi-squared density function. Each of the green dotted lines represents best achievable variance (Cram\'er Rao lower bound) of regression under a certain noise level.}
	\label{fig:sim_dist}
\end{figure*}

In the discussions above, the noise is assumed to have a single standard deviation. It is worthwhile to investigate the condition when the noise fluctuates and takes variable values to reveal the multimodal adaptability. For this purpose, we add noise with the original value and the three-fold value of standard deviation at equal probabilities. The standard deviation only varies \emph{between} examples; for a single example, the noise is kept unchanged.

In the second column of Fig. \ref{fig:sim_norm_calibre_uncorr} and Fig. \ref{fig:sim_norm_calibre_corr}, it can be seen that multi-modality does not change the performance of studied methods in a noteworthy
way. When the noise is uncorrelated, curve fitting is slightly better; when the noise is correlated, ensemble of neural networks is significantly better. Similar conclusions can be drawn from Table \ref{tab:sim_num}.

Fig. \ref{fig:sim_dist} illustrates multimodal distributions of predictive variance when the standard deviation of noise takes two distinct values. We also annotate the Cram\'er Rao lower bound \cite{ai2021neural} for each value of standard deviation. When the noise is uncorrelated, the distributions of predictive variance show good accordance to two modalities, while curve fitting is slightly better in the time section. When the noise is correlated, ensemble of neural networks performs steadily, while curve fitting tends to underestimate the larger variance especially in the energy section. This demonstrates that ensemble of neural networks can adapt to multi-modality as well as correlation.

\section{Experiment}
\label{sec:exp}

\begin{figure*}[!htb]
	\centering
	\includegraphics[width=.8\hsize]{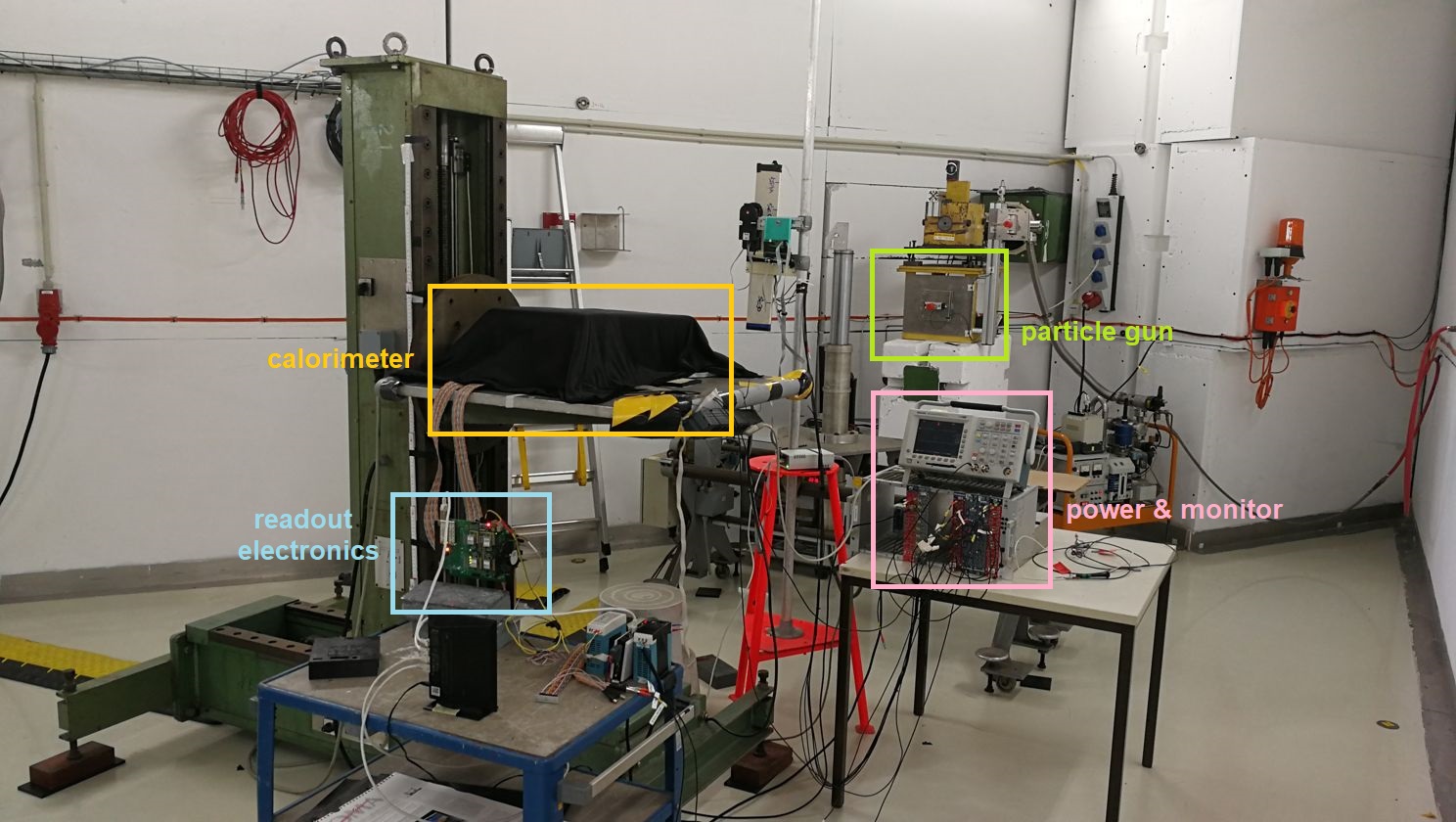}
	\caption{A photograph of the test beam scene of the NICA-MPD electromagnetic calorimeter at DESY.}
	\label{fig:detector_photo}
\end{figure*}

In this section, a 1 \si{GeV} electron test beam at DESY has been used to study the characteristics of an ECAL module. The detector response signal was recorded and digitized by a 5-\si{GSPS} 4-channel DRS4 data acquisition board \cite{DRS4EVAL}. Three channels were used to generate triggers and provide timing labels, and the other one was connected to the detector. This setting is the same as what was established in a former test beam study \cite{LI2020162833}. The test system is shown in Fig. \ref{fig:detector_photo}.

To train the neural networks, we construct a dataset of valid detector signals along with their timing labels and energy labels. The original waveform is sub-sampled with a ratio of 10:1 (500 \si{MSPS}). Here, the energy label is determined by measuring the pulse integral value before sub-sampling as in the simulation. This generates an energy precision (see Table \ref{tab:exp_num}) independent of the intrinsic energy resolution of the detector. Then the dataset is divided into the training set, comprising 8000 examples, and the test set, comprising 2000 examples. We train for 600 epochs with a batch size of 128, and another 300 epochs of quantization-aware training if quantization is used (in the last two rows of Table \ref{tab:exp_num}). Other configurations are as discussed in Section \ref{sec:sim_study}. In training, we find that the gradient descent optimization tends to overestimate the uncertainty because the inconsistency in training data prevents the model to stably find the optimum. This issue originates from the complex experimental environment and the limited model capacity. Hence, we apply bias-cancelling and sigma-scaling \cite{DBLP:conf/midl/LavesIFKO20} on the network outputs to improve the accuracy of the uncertainty estimation. For fair comparison, the same process is also applied to traditional methods.

\subsection{Predictive performance}

\begin{table*}[!htb]
	\centering
	\caption{Quantitative results of predictive performance for different conditions in the experiment. The bold number in each column of the table section (separated by research topics discussed below) indicates the best result of different methods/models.}
	\label{tab:exp_num}
	\small
	\begin{tabular}{c|ccc|ccc}
		\hline
		\multirow{2}{*}{} & \multicolumn{3}{c|}{time} & \multicolumn{3}{c}{energy} \\
		\cline{2-7}
		& NLL & precision (\si{ns}) & AUC & NLL & precision & AUC \\
		\hline
		baseline & \textbf{-5.965e-1} & \textbf{0.142} & 0.932 & \textbf{-4.064} & \textbf{0.60\%} & 0.950 \\
		dCFD \& int. & 1.382e & 0.878 & -- & 1.948 & 0.64\% & -- \\
		\hline
		baseline & \textbf{-5.965e-1} & \textbf{0.142} & 0.932 & -4.064 & 0.60\% & 0.950 \\
		3 conv. layers & -5.853e-1 & \textbf{0.142} & 0.932 & -4.460 & 0.41\% & 0.966 \\
		1 conv. layer & -5.546e-1 & 0.149 & 0.919 & \textbf{-4.775} & \textbf{0.29\%} & \textbf{0.985} \\
		2 fc layers  & -5.825e-1 & \textbf{0.142} & \textbf{0.940} & -4.138 & 0.55\% & 0.958 \\
		\hline
		baseline & \textbf{-5.965e-1} & \textbf{0.142} & 0.932 & \textbf{-4.064} & \textbf{0.60\%} & 0.950 \\
		8-bit quant. & -5.358e-1 & 0.149 & \textbf{0.936} & -3.749 & 0.89\% & \textbf{0.968} \\
		6-bit quant. & -4.801e-1 & 0.158 & 0.861 & -3.255 & 1.22\% & 0.905 \\
		\hline
	\end{tabular}
\end{table*}

\begin{figure*}[!htb]
	\centering
	\includegraphics[width=.75\hsize]{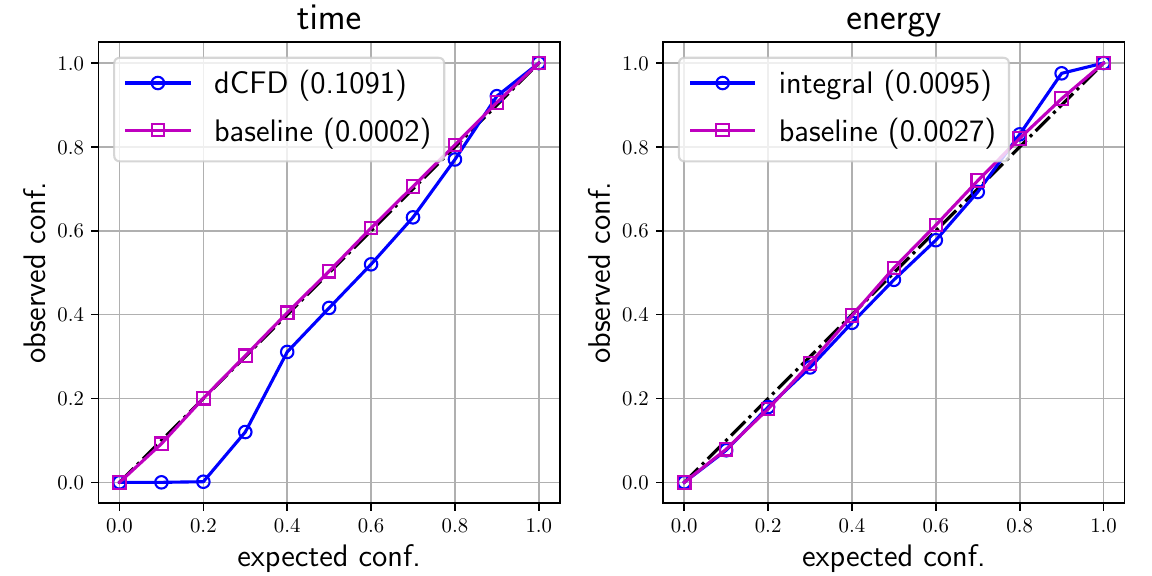}
	\caption{Normalized calibration plots for comparison of traditional methods and neural networks in the experiment. The numbers in the brackets are A-UCE scores.}
	\label{fig:exp_norm_calibre}
\end{figure*}

In Table \ref{tab:exp_num}, we list major test results in the experiment. The first two rows compare the baseline (ensemble of neural networks with architecture in Table \ref{tab:arch_exp}) and traditional methods, which use digital constant fraction discrimination (dCFD) \cite{FALLULABRUYERE2007247} for time measurement and waveform integration for energy measurement\footnote{For dCFD, we use half the maximum amplitude as the timing threshold and estimate the uncertainty by the slope crossing the threshold and noise level. For waveform integration, we integrate the waveform above the baseline and infer the uncertainty by charge and noise level.}. The energy precision of waveform integration is close to the baseline, while the timing precision of dCFD is much worse than the baseline. Besides, the estimated NLL values of traditional methods are much bigger than ensemble of neural networks. This is validated by Fig. \ref{fig:exp_norm_calibre}, where dCFD and waveform integration show a larger deviation from the straight line than the baseline. These demonstrate the limitation of traditional fixed algorithms being unable to characterize uncertainty precisely with unknown mathematical models.

In the next seven rows of Table \ref{tab:exp_num}, we list test results using ensemble of neural networks with different configurations. In all conditions except 6-bit quantization, the timing precision achieves better than 150 \si{ps}, and the best figure is 142 \si{ps}, which is significantly less than the value of 212.4 \si{ps} reported in \cite{LI2020162833}. Also in most cases, the energy precision is trivial compared to the reported 4.5\% energy resolution, which means that using ensemble of neural networks will not cause significant resolution loss for energy. These results confirm the outstanding accuracy of the proposed method when applied to experimental detector signals.

\subsection{Out-of-distribution detection}
\label{sec:out-of-dist}

\begin{figure*}[!htb]
	\centering
	\includegraphics[width=.75\hsize]{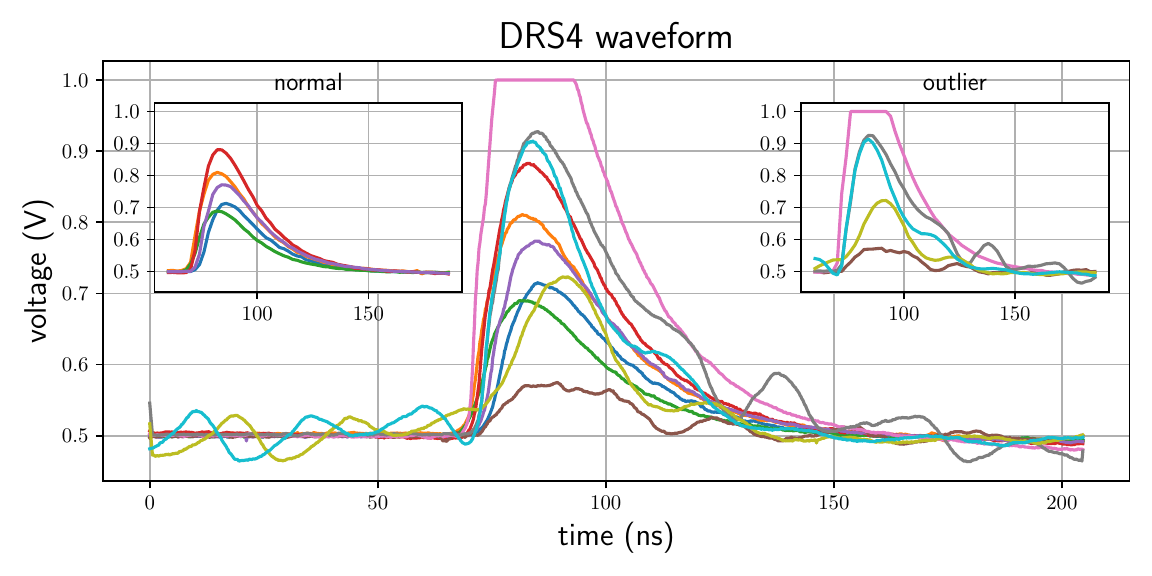}
	\caption{Examples of normal and outlying waveform observed in the experiment. The inset axes are waveform segments used to feed neural networks in the test stage.}
	\label{fig:wave_compare}
\end{figure*}

\begin{figure*}[!htb]
	\centering
	\includegraphics[width=.75\hsize]{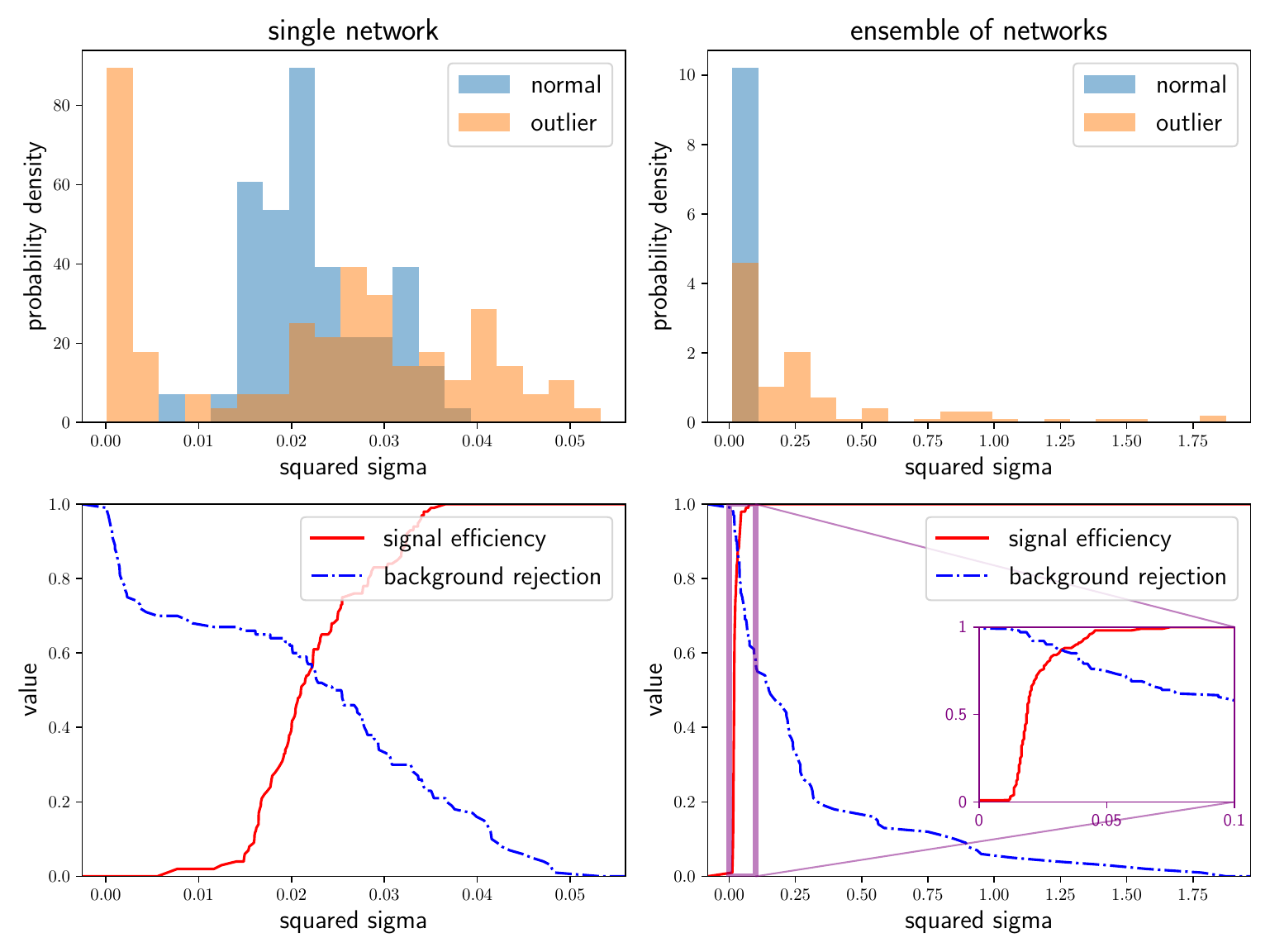}
	\caption{An illustration to display the different behaviors of a single neural network (left) and ensemble of networks (right) for out-of-distribution detection. We use the baseline condition and the time section (see Table \ref{tab:exp_num}) as an example to plot the figures.}
	\label{fig:exp_roc_explain}
\end{figure*}

\begin{figure*}[!htb]
	\centering
	\subfigure[]{
		\label{fig:exp_roc_hyper}
		\includegraphics[width=.75\hsize]{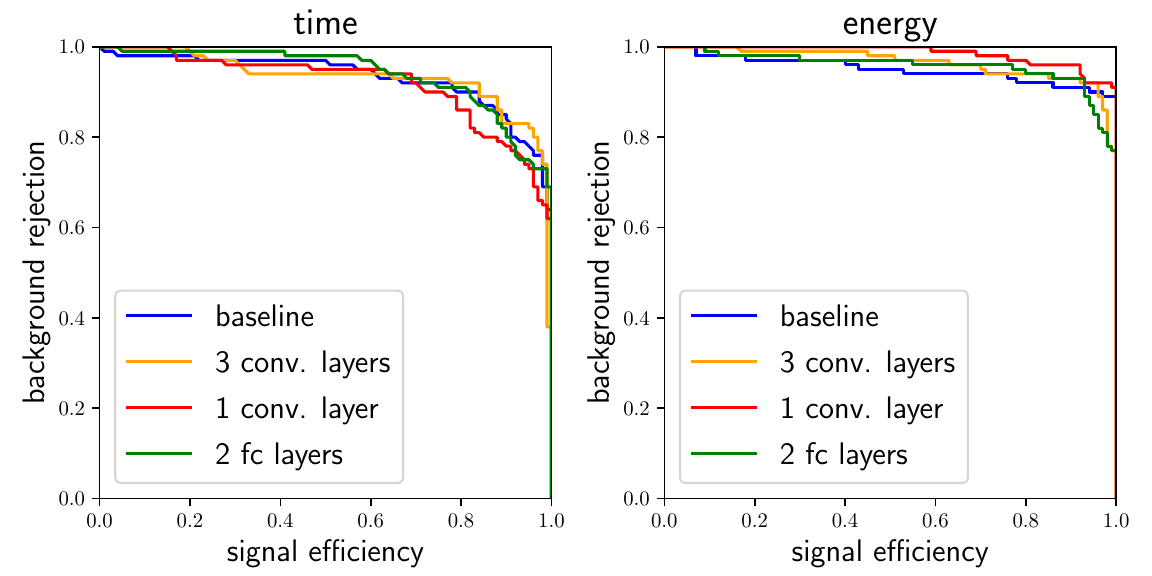}
	}
	\subfigure[]{
		\label{fig:exp_roc_quant}
		\includegraphics[width=.75\hsize]{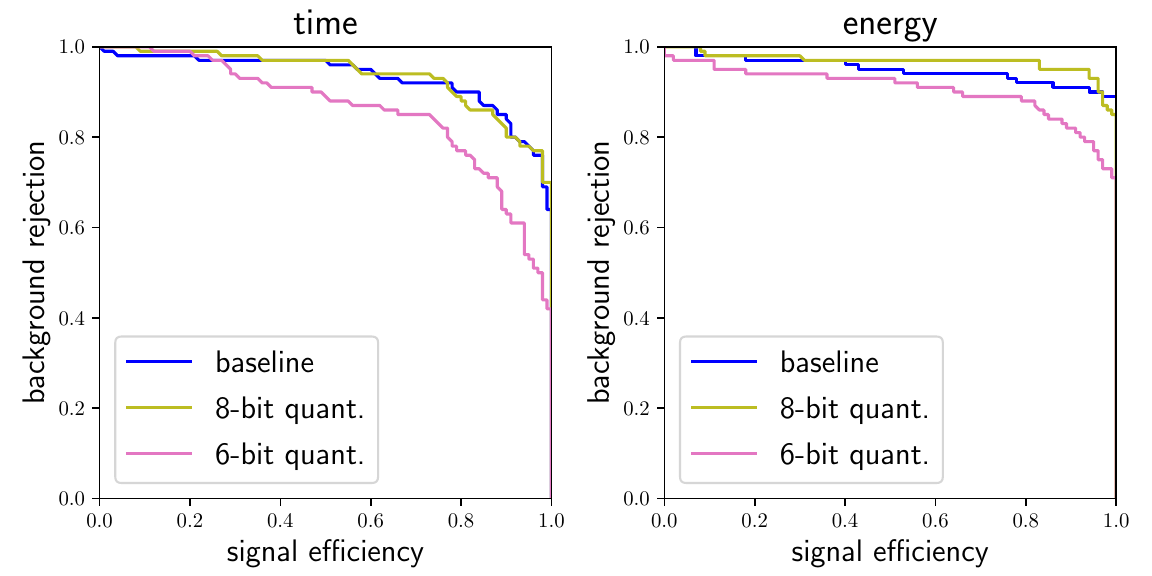}
	}
	\caption{(a) Receiver operating characteristic (ROC) curves for several network hyper-parameters based on the predictive variance of time and energy. (b) ROC curves for several quantization schemes based on the predictive variance of time and energy.}
\end{figure*}

Uncertainty estimation has many potential uses in high energy physics systems. For example, it can be propagated along with measurements for error analysis of specific physical targets. Here we focus on out-of-distribution detection, a very noticeable and visible enhancement to detect pulses with an abnormal signal shape. By setting a threshold for predictive variance, new signals unlike training data will be distinguished. No explicit statistical assumptions are needed in the process, and all knowledge about the data distribution is implicitly inherited in the neural network model. Unlike pulse shape discrimination which uses the exact number of classes (both normal signals and outliers) to train the model, here only examples passing a pre-selection stage are used in the training stage; in other words, our model is \emph{agnostic} to outliers (regarding their data distributions).

In Fig. \ref{fig:wave_compare}, we show some examples of normal and outlying waveform. By inspecting the outliers, they are either (i) overflowing above the dynamic range; (ii) displaying distinct vibration; or (iii) too small in amplitude. The normal signals, used to train the neural networks, show good uniformity and little deformation. When an outlying waveform is fed into the neural network model, it is expected to generate large predictive variance because of difficulties in determining precise predictions. Fig. \ref{fig:exp_roc_explain} illustrates the different behaviors between a single network and ensemble of networks with regard to the predictive variance. If we set a variance threshold, the signal efficiency can be defined as the ratio of normal signals below the threshold, and the background rejection as the ratio of outliers above the threshold. Good discriminators are able to keep both high at a certain threshold. It can be seen that, for the single network, the predictive variances of outliers scatter around normal signals, which is not desirable. On the other hand, ensemble of networks can push predictive variances of outliers well above the normal signals, and better discrimination is achieved. This strongly demonstrates that ensemble of neural networks is able to predict the epistemic uncertainty related to the distribution generating the data.

In receiver operating characteristic (ROC) curves, we change the variance threshold to estimate the background rejection vs. signal efficiency. The ROC curves by time/energy criteria are shown in Fig. \ref{fig:exp_roc_hyper} and Fig. \ref{fig:exp_roc_quant}. It can be seen in the ROC curves that the proposed method achieves satisfactory separation between normal signals and outliers. All curves are well above the neutral straight line from (0, 1) to (1, 0), indicating distinct differences in variance distributions and small overlapping regions. The area under the curve (AUC) is a numerical index to judge an ROC curve. In Table \ref{tab:exp_num}, most AUC values are above 0.9 (with only one exception). Besides, AUC values by energy criterion are better than corresponding AUC values by time criterion. This is reasonable, because changes in amplitude are more distinguishable than changes in timing and the predictive variance of energy will use them to give better estimates.

\subsection{Hyper-parameter sensitivity}
\label{sec:hyper_param_sens}

\begin{figure*}[!htb]
	\centering
	\includegraphics[width=.95\hsize]{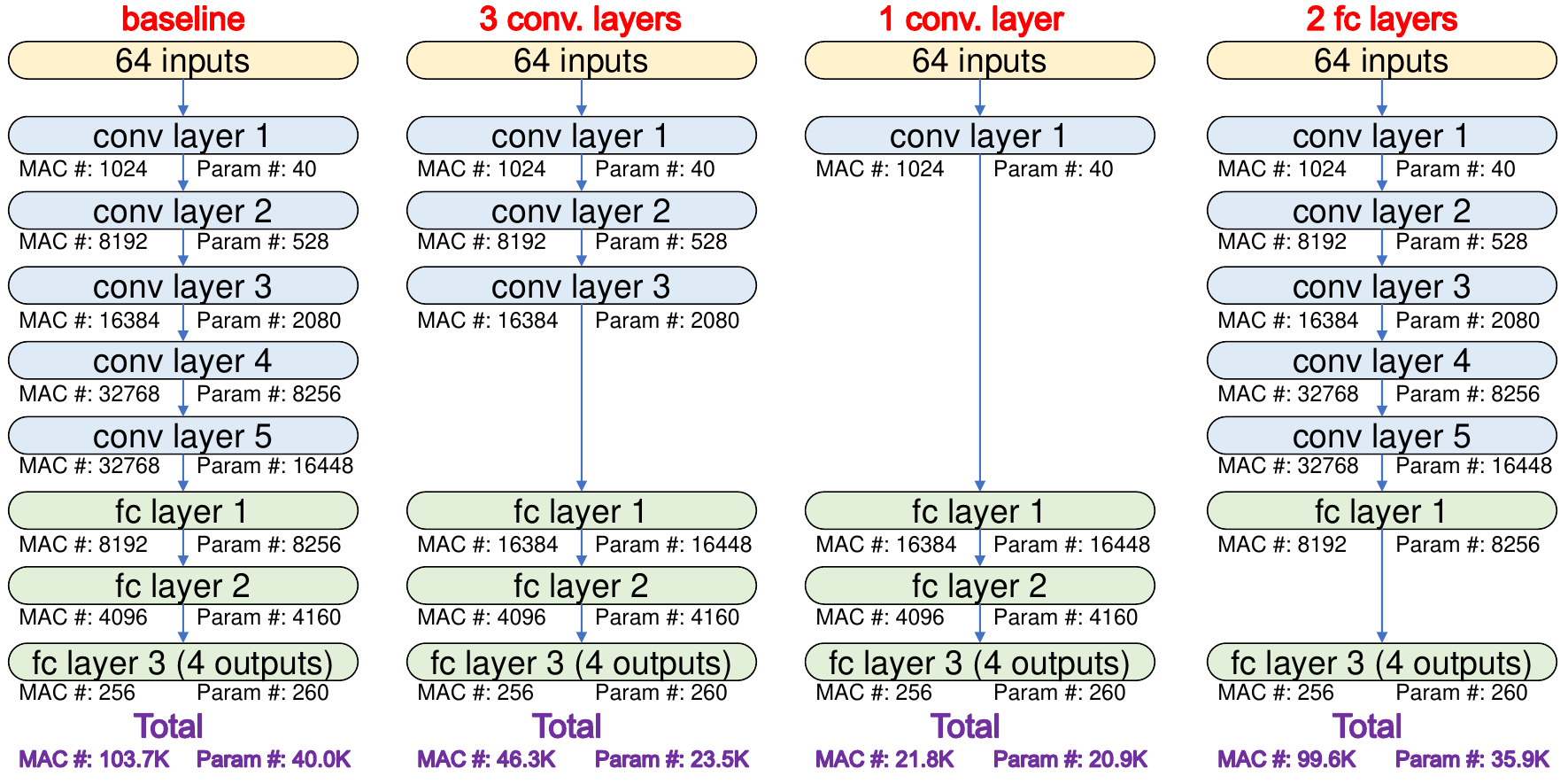}
	\caption{Hyper-parameter settings of neural networks used in the experiment. The amounts of multiply-and-accumulate (MAC) operations and trainable parameters for each layer and the whole network are annotated in this figure.}
	\label{fig:hyper_setting}
\end{figure*}

Neural networks are compute-intensive models. To customize the network architecture, it is worthwhile to explore different hyper-parameter settings with an emphasis on their impact on performance. We compare three more compact models with the baseline: reducing convolution layers to the upper three, reducing convolution layers to the upper one, and reducing fully-connected layers to two (the middle layer is removed). These model variants with associated MACs and trainable parameters are shown in Fig. \ref{fig:hyper_setting}. In the middle four rows of Table \ref{tab:exp_num} and also Fig. \ref{fig:exp_roc_hyper}, it can be seen that these compact models do not suffer obvious performance degradation. Actually, in the energy section, compact models exhibit some advantage with the experimental dataset being used (the best case is 1 convolution layer). In general, removing redundant layers will restrict the hypothesis space and alleviate the problem of over-fitting. However, one should bear in mind that dedicated structure and enough model capacity are essential to work in more complicated situations. In \cite{Ai_2019}, it was demonstrated that a fully-connected feed-forward neural network is unable to predict time in intense random Gaussian noise. Usually timing prediction is more challenging than energy prediction, and a well-structured network architecture is vital for network convergence in harsh situations.

\subsection{Quantization effect}
\label{sec:quan}

\begin{figure*}[!htb]
	\centering
	\includegraphics[width=.86\hsize]{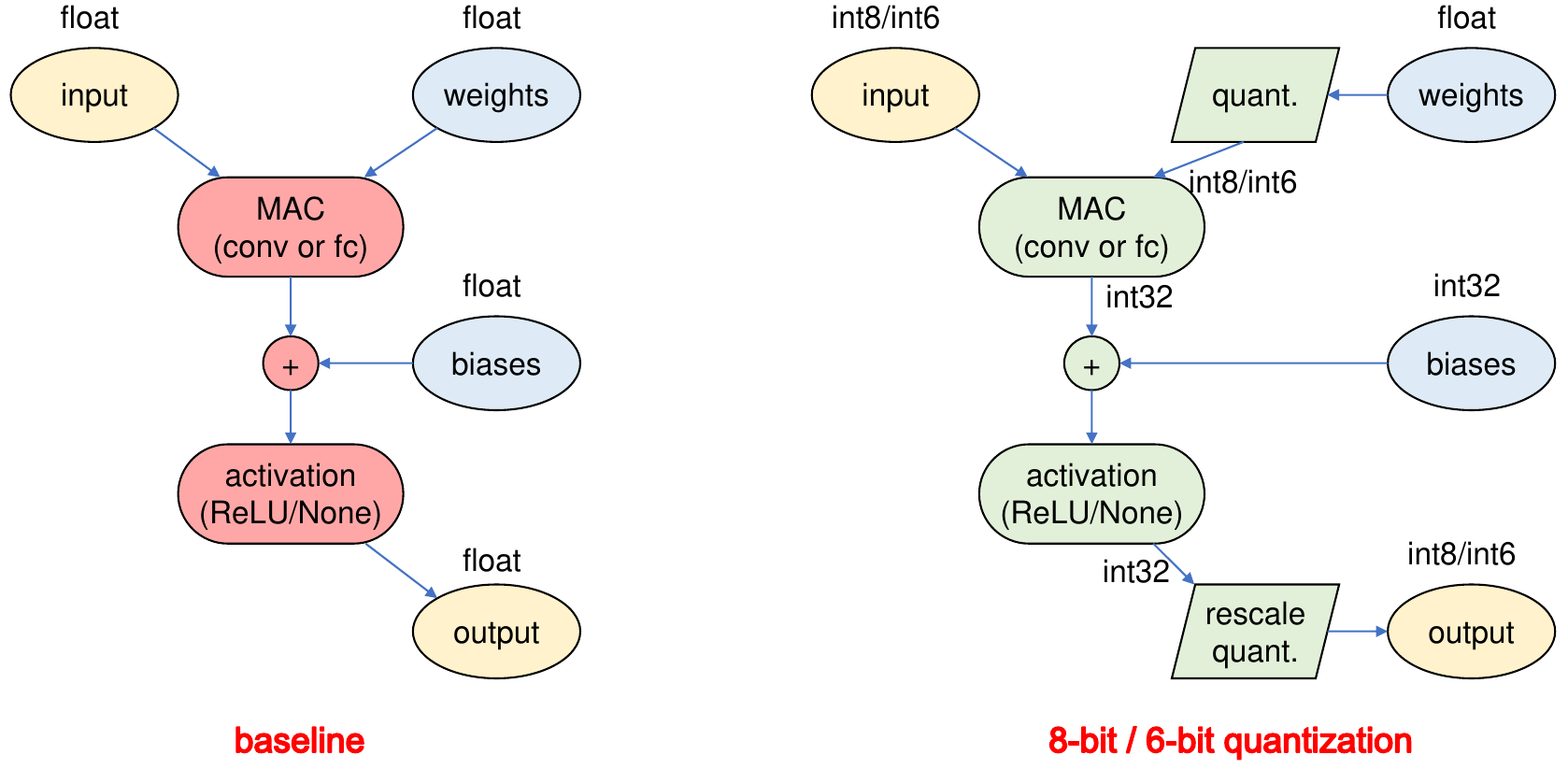}
	\caption{An illustration to compare the floating point operation (red) and the fixed point operation (green) after quantization in the neural network.}
	\label{fig:quant_flow}
\end{figure*}

It has been demonstrated above that the proposed method can work well as a computer software when waveform samples are present. Beyond being used offline, deploying the algorithm on front-end electronics is appealing because feature extraction can reduce the transmitted data substantially and thus decrease power consumption in the nuclear detector dataflow. For digital integrated circuits, it is much more economical to use low bit-width fixed point operations than floating point operations commonly used in computer software. To evaluate the effect of quantization, we quantize the weights and activations of the network to 8-bit fixed point and 6-bit fixed point, respectively. The procedure of quantization and fixed point operation is shown in Fig. \ref{fig:quant_flow}. The test results are shown in the bottom three rows of Table \ref{tab:exp_num} and also Fig. \ref{fig:exp_roc_quant}. In summary, using 8-bit quantization will shrink the performance slightly and the degradation is totally acceptable; using 6-bit quantization is more aggressive with a significant performance degradation. One should judge the most appropriate quantization scheme as a balance of cost and accuracy.

\section{Conclusion}

In this paper, a novel method based on data science and artificial intelligence, namely ensemble of neural networks, is proposed for the uncertainty estimation of nuclear detector signals. The network architecture is specially designed to produce the predictive mean and predictive variance of physics-related features with the help of a dedicated loss function. Specifically, the proposed method is an extension to \cite{DBLP:conf/nips/Lakshminarayanan17} and \cite{DBLP:conf/nips/KendallG17} in the nuclear detector area. Both simulations and experiments prove that the method provides satisfactory predictions of the signal features, as well as a justifiable uncertainty of those predictions. The application of the method is multi-faceted and far-reaching. We hope this work will benefit the community for the conceptualization and design of future detector systems in physical experiments.

\acknowledgments

This research is supported by the National Key Research and Development Program of China (under project no. 2020YFE0202001). This research is also supported by China Postdoctoral Science Foundation (under grant no. 2021M690088).

\appendix

\section{Performance metrics and normalized calibration plot}

\begin{algorithm}[H]
	\centering
	\caption{Computation of performance metrics and normalized calibration plot (NCP).}
	\label{alg:metrics}
	\small
	\begin{algorithmic}
		\State {\bfseries Input:} data: $\bm{x}^{(1)}, \bm{x}^{(2)}, ..., \bm{x}^{(N)} \in \mathbb{R}^L$, ground-truth label: $\bm{y}^{(1)}, \bm{y}^{(2)}, ..., \bm{y}^{(N)} \in \mathbb{R}^M$, predictive model: $f: \mathbb{R}^L \rightarrow \mathbb{R}^{2M}$
		\State {\bfseries Input:} normal density: $p(x;\mu,\sigma)$, normal distribution: $F(x;\mu,\sigma)$, number of uncertainty levels: $N_B, N_A$
		\State {\bfseries Output:} NLL, B-UCE, A-UCE, bias, precision and NCP
		\For{$i=1$ {\bfseries to} $N$}
		\State propagate $\bm{x}^{(i)}$ throughout the model: $(\bm{\mu}^{(i)}, {\bm{\sigma}^2}^{(i)}) \gets f(\bm{x}^{(i)})$, where $\bm{\mu}^{(i)}, {\bm{\sigma}^2}^{(i)} \in \mathbb{R}^M$
		\EndFor
		\For{$j=1$ {\bfseries to} $M$}
		\State $\text{NLL}_j \gets - \frac{1}{N} \sum\limits_{i=1}^{N} \log p(y_j^{(i)};\mu_j^{(i)},\sigma_j^{(i)}) $
		\State $\breve{\sigma}^2_j \gets \max\limits_{i \in \{ 1, ..., N \}} {\sigma^2_j}^{(i)}$ and $\check{\sigma}^2_j \gets \min\limits_{i \in \{ 1, ..., N \}} {\sigma^2_j}^{(i)}$
		\State $\mathcal{U}_k \gets \left\{ i\ |\ (\breve{\sigma}^2_j - \check{\sigma}^2_j) \cdot \frac{(k-1)}{N_B} + \check{\sigma}^2_j \leq {\sigma^2_j}^{(i)} < (\breve{\sigma}^2_j - \check{\sigma}^2_j) \cdot \frac{k}{N_B} + \check{\sigma}^2_j \right\}$ for $k = 1, 2, ..., N_B$
		\State $\text{B-UCE}_j \gets \frac{1}{N}\sum\limits_{k=1}^{N_B} \left| \sum\limits_{i \in \mathcal{U}_k} || y_j^{(i)} - \mu_j^{(i)} ||^2 - \sum\limits_{i \in \mathcal{U}_k}{\sigma^2_j}^{(i)} \right|$
		\State $\mathcal{V}_k \gets \left\{ i\ |\ F(y_j^{(i)};\mu_j^{(i)},\sigma_j^{(i)}) < \frac{k}{N_A} \right\}$ for $k = 0, 1, 2, ..., N_A$
		\State $\text{A-UCE}_j \gets \sum\limits_{k=0}^{N_A} \left( \frac{k}{N_A} - \frac{|\mathcal{V}_k|}{N} \right)^2$ and plot $\{(\frac{k}{N_A}, \frac{|\mathcal{V}_k|}{N})\}_{k=0}^{N_A}$ as $\text{NCP}_j$
		\State $\text{bias}_j \gets \frac{1}{N} \sum\limits_{i=1}^{N} \left( \mu_j^{(i)} - y_j^{(i)} \right)$
		\State $\text{precision}_j \gets \sqrt{ \frac{1}{N} \sum\limits_{i=1}^{N} \left( \mu_j^{(i)} - y_j^{(i)} - \text{bias}_j \right)^2 }$
		\EndFor
	\end{algorithmic}
\end{algorithm}

\section{Individual neural networks in the ensemble}

\begin{figure}[H]
	\centering
	\includegraphics[width=.65\hsize]{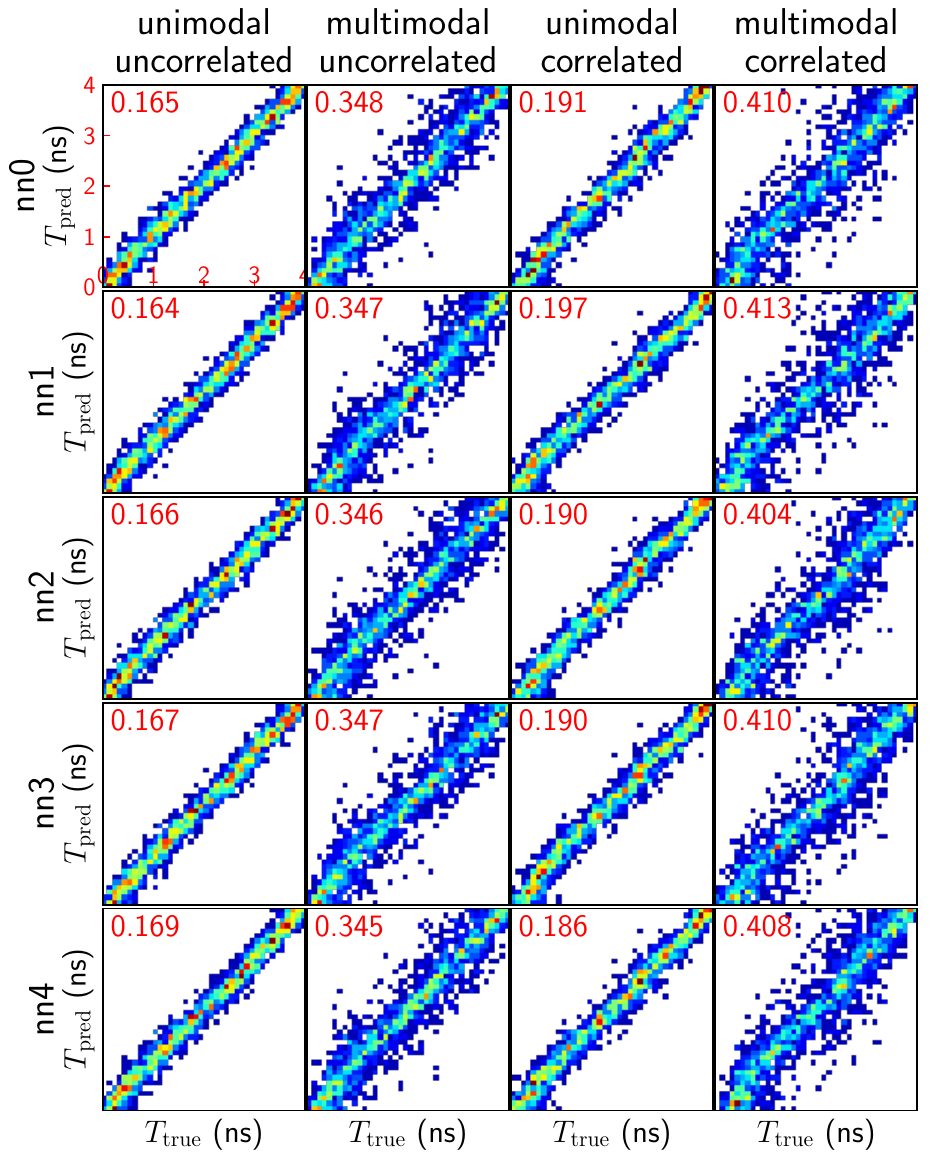}
	\includegraphics[width=.65\hsize]{sim_pred_value_cb_renew.pdf}
	\caption{Two-dimensional histograms show predicted time vs. ground-truth time using individual neural networks in four conditions. The red numbers in the top left-hand corner of each box indicate timing precision in \si{ns}. The axis scales are identical in each case, and are shown in the top left-hand box.}
	\label{fig:sim_pred_value_supp}
\end{figure}


\providecommand{\href}[2]{#2}\begingroup\raggedright\endgroup

\end{document}